\documentclass[reprint,prl,aps,notitlepage,amsmath,amssymb,superscriptaddress,nobibnotes]{revtex4-1}
\usepackage{graphicx}
\usepackage{bm}
\usepackage{xr}
\usepackage{color}

\usepackage[compact]{titlesec}
\titlespacing\section{0pt}{4pt plus 2pt minus 2pt}{0pt plus 2pt minus 2pt}
\titlespacing\subsection{0pt}{4pt plus 2pt minus 2pt}{0pt plus 2pt minus 2pt}
\titlespacing\subsubsection{0pt}{4pt plus 2pt minus 2pt}{0pt plus 2pt minus 2pt}

\begin{document}
\title{Orbital symmetry of charge density wave order in La$_{1.875}$Ba$_{0.125}$CuO$_4$ and YBa$_2$Cu$_3$O$_{6.67}$}

\author{A. J. Achkar}
\affiliation{Department of Physics and Astronomy, University of Waterloo, Waterloo, N2L 3G1, Canada}
\author{F. He}
\affiliation{Canadian Light Source, Saskatoon, Saskatchewan, S7N 2V3, Canada}
\author{R. Sutarto}
\affiliation{Canadian Light Source, Saskatoon, Saskatchewan, S7N 2V3, Canada}
\author{Christopher McMahon}
\affiliation{Department of Physics and Astronomy, University of Waterloo, Waterloo, N2L 3G1, Canada}
\author{M. Zwiebler}
\affiliation{Leibniz Institute for Solid State and Materials Research IFW Dresden, Helmholtzstra{\ss}e 20, 01069 Dresden, Germany}
\author{M. H\"{u}cker}
\affiliation{Condensed Matter Physics and Materials Science Department, Brookhaven National Laboratory, Upton, NY 11973, USA}
\author{G. D. Gu}
\affiliation{Condensed Matter Physics and Materials Science Department, Brookhaven National Laboratory, Upton, NY 11973, USA}
\author{Ruixing Liang}
\affiliation{Department of Physics and Astronomy, University of British Columbia, Vancouver,V6T 1Z1, Canada}
\affiliation{Canadian Institute for Advanced Research, Toronto, Ontario M5G 1Z8, Canada}
\author{D. A. Bonn}
\affiliation{Department of Physics and Astronomy, University of British Columbia, Vancouver,V6T 1Z1, Canada}
\affiliation{Canadian Institute for Advanced Research, Toronto, Ontario M5G 1Z8, Canada}
\author{W. N. Hardy}
\affiliation{Department of Physics and Astronomy, University of British Columbia, Vancouver,V6T 1Z1, Canada}
\affiliation{Canadian Institute for Advanced Research, Toronto, Ontario M5G 1Z8, Canada}
\author{J. Geck}
\affiliation{Chemistry and Physics of Materials, Paris Lodron University Salzburg, Hellbrunner Strasse 34, 5020 Salzburg, Austria}
\author{D. G. Hawthorn}
\affiliation{Department of Physics and Astronomy, University of Waterloo, Waterloo, N2L 3G1, Canada}
\affiliation{Canadian Institute for Advanced Research, Toronto, Ontario M5G 1Z8, Canada}

\begin{abstract}
Recent theories of charge density wave (CDW) order in high temperature superconductors have predicted a primarily 
$d$ CDW orbital symmetry. Here, we report on the orbital symmetry of CDW order in the canonical cuprate 
superconductors La$_{1.875}$Ba$_{0.125}$CuO$_4$ (LBCO) and YBa$_2$Cu$_3$O$_{6.67}$ (YBCO), using 
resonant soft x-ray scattering and a model mapped to the CDW orbital symmetry. From measurements sensitive to 
the O sublattice, we conclude that LBCO has predominantly $s'$ CDW orbital symmetry, in contrast to the $d$ orbital 
symmetry recently reported in other cuprates. Additionally, we show for YBCO that the CDW orbital symmetry differs along the $a$ and $b$ crystal axes and that these both differ from LBCO. This work highlights CDW orbital symmetry as an additional key property that distinguishes the different cuprate families. We discuss how the CDW symmetry may be related to the ``1/8--anomaly'' and to static spin ordering.
\end{abstract}


\date{\today}
\maketitle

Charge density wave (CDW) order in underdoped cuprates has recently been revealed as an important and generic competitor to superconductivity (SC).\cite{Tranquada95,Kohsaka07,Wu11,Ghiringhelli12,Chang12, Achkar12,Blackburn13,Comin14a,daSilvaNeto14} A significant property of CDW order is that it can exhibit both {\it inter} and {\it intra} unit cell symmetry breaking.\cite{Vojta09,Lawler10}  Specifically, CDW order can occur with $d$ rather than $s$ or $s'$ orbital symmetry.\cite{Li06b,Seo07,Vojta08,Metlitski10,Sachdev13,Efetov13,Atkinson14,Fujita14,Allais14,Chowdhury14b} Here, we report resonant soft x-ray scattering (RSXS) measurements of La$_{1.875}$Ba$_{0.125}$CuO$_4$ (LBCO) and YBa$_2$Cu$_3$O$_{6.67}$ (YBCO) that are resolved onto the O 2$p_{x(y)}$ and Cu sublattices and mapped to the CDW symmetry. Our main finding is that CDW order in LBCO has primarily $s'$ symmetry with a secondary $d$ component, distinguishing it from the predominant $d$ symmetry CDW in Bi$_2$Sr$_2$CaCu$_2$O$_{8+\delta}$ (Bi-2212) and Ca$_{2-x}$Na$_x$CuO$_2$Cl$_2$ (Na-CCOC),\cite{Fujita14}  and YBCO.\cite{Comin14c} We propose that the $s'$ symmetry in LBCO may be related to the ``1/8--anomaly'' and that it favours static spin ordering more than $d$ symmetry. Additionally, we find that the CDW orbital symmetry in YBCO differs along the $a$ and $b$ crystal axes, indicating CDW order with unidirectional character. Finally, we present and discuss implications of energy dependent scattering from the O 2$p_{x(y)}$ sublattices in LBCO. 

Stimulated by theory,\cite{Sachdev13,Efetov13} evidence for a mixed $d+sÕ$ symmetry CDW with prominent $d$ component was reported in YBCO by RSXS at the Cu $L$ edge,\cite{Comin14c} and a dominant $d$ symmetry character to the CDW was revealed by STM in Bi-2212 and Na-CCOC.\cite{Fujita14} In the $d$-symmetry CDW state, the modulation of charge (or a related microscopic quantity) on O $p_x$ and O $p_y$ sites is out of phase. An important question is whether $d$-symmetry CDW order is a generic property of underdoped cuprates and, specifically, if it also occurs in the canonical stripe-ordered La-based cuprates. There are many similarities in the CDW order of the La-based cuprates and other cuprates (eg. Bi-2212, YBCO) such as an enhancement in CDW intensity at doping levels near $p=1/8$,\cite{Hucker11,Fujita14a,Huecker14,BlancoCanosa14} competition with SC and a common spectroscopic signature to the resonant scattering intensity.\cite{Achkar12,Achkar13} However, these similarities are at odds with important differences such as the doping dependence of the CDW incommensurability.\cite{Hucker11,Blackburn13,Huecker14,BlancoCanosa14,Fujita12,Wise08,Yamada98,Comin14a} Perhaps most significantly, static spin density wave (SDW) order commensurate with CDW order is only observed in the La-based cuprates. Accordingly, it is not yet clear whether stripe order in the La-based cuprates and CDW order in other cuprates are slightly different manifestations of a common order or truly distinct phases. 

\begin{figure}
\centering
\resizebox{\columnwidth}{!}{\includegraphics{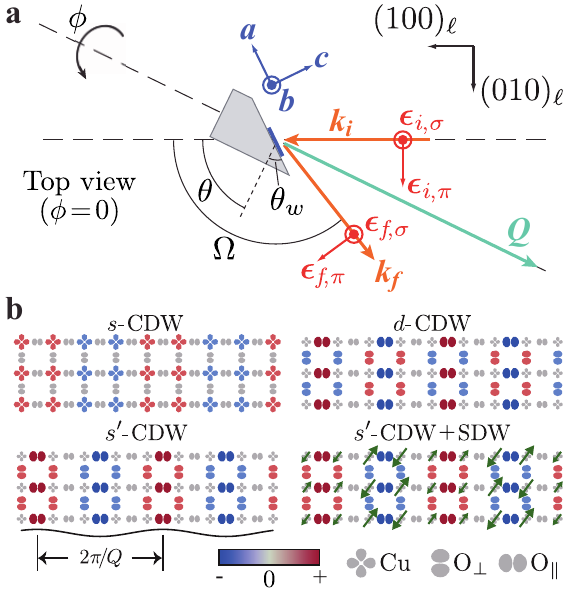}}
\caption{{\bf Scattering geometry and schematics of CDW order.} {\bf a.} The experimental setup in the laboratory frame, $\ell$. The scattering angles are defined by the sample rotation $\theta$, the wedge angle $\theta_w$, the detector angle $\Omega$ and the azimuthal rotation angle, $\phi$. $\theta_w$ is selected to align the $\phi$ and $\bm{Q}$ axes, allowing the CDW order to be continually probed while rotating the scattering tensor. {\bf b.} CDW order having $s$, $s'$ and $d$ orbital symmetries in the CuO$_2$ plane are depicted for a bond-centered, commensurate, and unidirectional CDW. Color denotes the modulation of the charge from the average, as defined in Supplementary equation (S13) by $\Delta_{ij}$. Green arrows are spins.}
\label{fig:ScatGeom_Models}
\end{figure}

Here, we have resolved the orbital symmetry and microscopic character of CDW order in LBCO and YBCO using a RSXS technique where the incident photon polarization is varied relative to the material's crystallographic axes and the crystal is simultaneously rotated about the CDW ordering wavevector $\bm{Q}$,\cite{Comin14c} as illustrated in Fig.~\ref{fig:ScatGeom_Models}a. With this approach, we are able to determine the relative strength, phase relation and energy dependence of scattering from two O sublattices comprised of O atoms with Cu--O--Cu bonds either parallel (O$_\parallel$) or perpendicular (O$_\perp$) to $\bm{Q}$ (see Fig.~\ref{fig:ScatGeom_Models}b). The proportion of $d$ to $s'$ symmetry was characterized from O sublattice measurements of LBCO using a polarization dependent RSXS model developed with parameters directly related to symmetry components $\Delta_d$ and $\Delta_{s'}$ defined by theory.\cite{Sachdev13} Additionally, RSXS measurements from the Cu sublattices of LBCO and YBCO identified different components to their scattering tensors, and accordingly different symmetry proportions, not only between compounds but also for CDW order along $a$ and $b$ in YBCO. 

Measurements of resonant x-ray scattering were performed at photon energies corresponding to the Cu $L$ ($2p\to 3d$, 931.4 eV) and O $K$ ($1s\to 2p$, 528.3 eV) x-ray absorption edges (see Methods). At these energies, the CDW order can be resolved into Cu and O sublattices, which occupy the ``sites'' and ``bonds'', respectively, of the CuO$_2$ plane. Since O 2$p$ holes are primarily in $\sigma$-bonded $2p_x$ or $2p_y$ orbitals, the O sublattice can be further subdivided into two sublattices, O$_\parallel$ and O$_\perp$ that are $\sigma$-bonded parallel or perpendicular, respectively, to the in-plane component of the CDW wavevector, $\bm{Q}$ (see Fig.~\ref{fig:ScatGeom_Models}b).  Although the average orbital occupation may be the same for O$_\parallel$ and O$_\perp$, at the O $K$ edge, how strongly x-rays scatter from O$_\parallel$ relative to O$_\perp$  will depend strongly upon photon polarization, $\bm{\epsilon}$, and sample geometry, which can be varied by rotating the sample about an azimuthal angle, $\phi$ at fixed $\bm Q$ or by varying the $L$ component of $\bm{Q}$. 

Specifically, as shown in the Supplementary Information, the polarization/geometry dependent scattering intensity provides a measure of the ratio ${t_\parallel}/{t_\perp}$:
\begin{equation}
\frac{t_\parallel}{t_\perp}=\frac{\sum_n f_{\text{O}_\parallel,n} e^{-i \bm{Q}\cdot \bm{r}_n}}{\sum_n f_{\text{O}_\perp,n} e^{-i \bm{Q}\cdot \bm{r}_n}},
\label{eqn:tpartperp}
\end{equation}
where $t_{\parallel}$ ($t_{\perp}$) is a structure factor for the component of O 2$p$ orbitals projected onto a direction within the CuO$_2$ plane and parallel (perpendicular) to the CDW wavevector, $\bm{Q}$.  Here, $f_{\text{O}_{\parallel (\perp)},n}$ is the atomic scattering form factor for O$_{\parallel (\perp)}$ and $n$ is the site index.  A key aspect of $t_\parallel/t_\perp$ is that it provides sensitivity to both the magnitude and relative phase of the modulations on the O$_\parallel$ and O$_\perp$ sublattices.  Note, we also measure  $t_{cc}/t_\perp$, where $t_{cc}$ corresponds to a structure factor for the component of O 2$p$ orbitals projected onto the $c$-axis.

For the O sublattice, $t_\parallel/t_\perp$ is particularly relevant because it can be directly mapped to the orbital symmetry of the CDW order.  As shown in the Supplementary Information, we use the model of Ref.~\cite{Sachdev13} to relate $t_\parallel/t_\perp$ to the relative amplitude of the $d$ and $s'$ symmetry components of the CDW order, $\Delta_{d}/\Delta_{s'}$:
\begin{equation}
\frac{\Delta_d}{\Delta_{s'}}=\frac{t_\parallel/t_\perp-1}{t_\parallel/t_\perp+1},
\label{eqn5}
\end{equation} Inspection of equation~(\ref{eqn5}) shows that for pure $d$-CDW ($s'$-CDW) order, $t_{\parallel}/t_{\perp}=-1$ ($t_{\parallel}/t_{\perp}=1$). A mixed $d$ and $s'$ state would have $|t_{\parallel}/t_{\perp}| \neq 1$.


Measurements of the Cu sublattice also provide a valuable measure of the CDW orbital symmetry, with $t_\parallel/t_\perp$ associated with the symmetry of modulations of the Cu $2p$ and $3d$ states (see Supplementary Information).  Being primarily sensitive to the Cu ``sites'', these measurements are not as directly mapped to $\Delta_d/\Delta_{s'}$, as discussed below (see Fig.~1{\bf b}). The CDW symmetry has been quantified, however, using Cu $L$ edge RSXS measurements in YBCO\cite{Comin14c} where a variation of the O $K$ edge modelling presented here is used to model the CDW symmetry in terms of parameters $\delta_{s,p,d}$ that are related to $\Delta_{s,p,d}$.


\begin{figure}
\centering
\resizebox{\columnwidth}{!}{\includegraphics{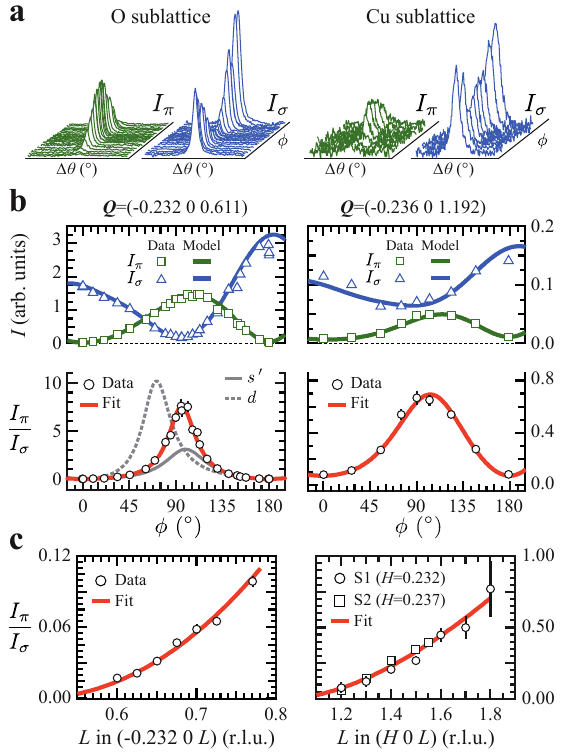}}
\caption{{\bf Resonant scattering from CDW order on the O and Cu sublattices of LBCO.}  {\bf a.} Azimuthal angle dependence of the background-subtracted RSXS signal with incident $\sigma$ and $\pi$ polarized light. {\bf b.} Top panels: Measured scattering amplitudes $I_\pi$ (${\color[rgb]{0.082309,0.346869,0.034636} \Box}$) and $I_\sigma$ (${\color[rgb]{0,0,1} \bigtriangleup}$). Lower panels: the ratio $I_\pi/I_\sigma$ ($\bigcirc$).  {\bf c.} The ratio $I_\pi/I_\sigma$ plotted against $L$. Simultaneously fitting the $I_\pi/I_\sigma$ data in {\bf b} and {\bf c} gives the red solid lines, corresponding to the $t_{\parallel}/t_{\perp}$ and $t_{cc}/t_{\perp}$ parameters reported in the text. The blue and green lines in {\bf b} are also calculated using these parameters. For the O sublattice, this corresponds to a CDW state with mixed $s'$ and $d$ symmetry, dominated by the $s'$ component. Calculations for pure $s'$ (solid gray) and $d$ (dashed gray) CDW states clearly fail to describe the experimental data. Error bars in panels {\bf b} and {\bf c} are from the 2$\sigma$ standard deviation in the peak amplitude from Lorentzian fits to $I_\pi$ and $I_\sigma$.}
\label{fig:LBCO_PolDep}
\end{figure}

In Fig.~\ref{fig:LBCO_PolDep}, we investigate the $\phi$ and $L$ dependence of RSXS from CDW order in 1/8-doped LBCO at the O $K$ (left column) and Cu $L$ (right column) x-ray absorption edges. The dependence on $\phi$ of $I_\pi/I_\sigma$ constrains both the magnitude and sign of $t_{\parallel}/t_{\perp}$ whereas the dependence on $L$ constrains only its magnitude. Note, the $L$ dependence is used here to probe the orbital symmetry of the CDW order by varying the sample photon polarization vs. relative to the crystal axes.  Accounting for these geometry dependence $I_\pi/I_\sigma$ can be fit with $L$-independent $t_{\parallel}/t_{\perp}$ and $t_{cc}/t_{\perp}$.  In Fig.~\ref{fig:LBCO_PolDep}a, the background-subtracted RSXS signal of the CDW order is shown for incident $\pi$ and $\sigma$ photon polarization. The $\phi$ dependence of the fit amplitudes and their ratio $I_\pi/I_\sigma$ are shown in Fig.~\ref{fig:LBCO_PolDep}b. The $L$ dependence of $I_\pi/I_\sigma$ is shown in Fig.~\ref{fig:LBCO_PolDep}c. This was measured in two samples, denoted S1 and S2, for the Cu sublattice.

For the O and Cu sublattices, simultaneously fitting $I_\pi/I_\sigma$ for both the $\phi$ (Fig.~\ref{fig:LBCO_PolDep}b, lower panel) and $L$ (Fig.~\ref{fig:LBCO_PolDep}c) dependence gives ($t_{\parallel}/ t_{\perp} = 0.612\pm 0.035$, $t_{cc}/t_{\perp} = 0.034\pm 0.021$) and ($t_{\parallel}/t_{\perp} =  0.991\pm0.015$, $t_{cc}/t_{\perp} = -0.067\pm 0.015$), respectively (see Supplementary Information for discussion of parameter estimation and confidence regions).

To illustrate the experimental and model sensitivity to $\Delta_d/\Delta_{s'}$ on the O sublattice, calculations for $t_{\parallel}/t_{\perp} =1$ ($t_{\parallel}/t_{\perp} =-1$), corresponding to pure $s'$ ($d$) symmetry CDW order, are shown in Fig.~\ref{fig:LBCO_PolDep}b in solid (dashed) gray. The calculated $I_\pi/I_\sigma$ curves in these pure symmetry states clearly fail to describe the experimental data. Furthermore, in Fig.~\ref{fig:LBCO_PolDep}b (top panels), we show that calculations (solid lines) of the $\phi$ dependence using the best fit parameters capture the measured $I_\pi$ (green squares) and $I_\sigma$ (blue triangles) very well (not just their ratio), providing confidence that all relevant factors such as surface geometry and absorption corrections have been accounted for. 

For the O sublattice, the ratio $t_{\parallel}/t_{\perp}\sim 0.6$ indicates that the scattering tensor in LBCO breaks four-fold rotational symmetry by inducing a smaller modulations on the O$_{\parallel}$ sublattice than the O$_{\perp}$ sublattice.  Interpreting $t_{\parallel}/t_{\perp}$ in terms of orbital symmetry using equation (\ref{eqn5}), we find that the CDW order in LBCO has mixed, anti-phase $d+s'$ symmetry with a dominant $s'$ symmetry: $\Delta_d/\Delta_{s'} = -0.241 \pm 0.027$.  This result stands in contrast to the dominant $d$ symmetry CDW reported in Bi-2212 and Na-CCOC by STM,\cite{Fujita14} and the prominent d-symmetry observed in Cu $L$ edge RSXS from YBCO.\cite{Comin14c}

It should be noted that the deviation of $t_{\parallel}/t_{\perp}$ from unity that we observed (for the O sublattice) is too large to result solely from the low-temperature tetragonal (LTT) phase in LBCO. The LTT distortion cants the CuO$_6$ octahedra making each CuO$_2$ plane orthorhombic and breaking the four-fold rotational symmetry of the lattice. However, the anisotropy in the average electronic structure induced by this distortion is expected to be small.  For instance, a 3\% anisotropy in hopping parallel and perpendicular to $\bm{Q}$,\cite{Kampf01} can be deduced from measured bond angles. The anisotropy in the average structure factors, and thus in $t_{\parallel}/t_{\perp}$, would be similar. 

\begin{figure}
\centering
\resizebox{\columnwidth}{!}{\includegraphics{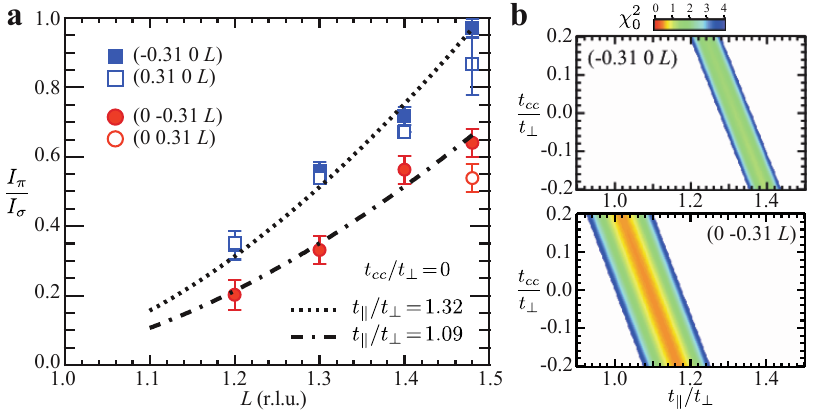}}
\caption{{\bf Cu sublattice scattering in YBCO}. {\bf a.} $L$ dependence of $I_\pi/I_\sigma$ for CDW order having $\bm{Q}=(\pm0.31 ~0 ~L)$ (blue) and $\bm{Q}=(0~\pm0.31 ~L)$ (red) with photon energy 931.3 eV. The peaks along $H$ have larger $I_\pi/I_\sigma$ than those along $K$, giving different asymmetries in $t_{\parallel}/t_{\perp}$ for the different directions. Fits to the data (dashed lines) are for $t_{cc}=0$. {\bf b.} Maps of $\chi_0^2$ show linear regions of acceptable fit parameters for $H$ (top) and $K$ (bottom).  Error bars in panel {\bf a} are the standard deviation in the peak amplitude from Lorentzian fits to $I_\pi$ and $I_\sigma$.}
\label{fig:YBCO_LDep} 
\end{figure}

At the Cu $L$ edge, we have investigated the CDW symmetry in both LBCO and YBCO.  In contrast to scattering that is nearly isotropic within the CuO$_2$ plane in LBCO ($t_{\parallel} \approx t_{\perp}$), $t_{\parallel}/t_{\perp}$ is more anisotropic in YBCO.  Specifically, from fits to the $L$ dependence of the CDW peaks in YBCO (Fig.~\ref{fig:YBCO_LDep}) give $1.01\leq|t_{\parallel}/t_{\perp}|\leq1.17$ for $\bm{Q} = (0 \pm ~0.31 ~L$) and $1.24\leq|t_{\parallel}/t_{\perp}|\leq1.39$ for $\bm{Q} =(\pm 0.31 ~0 ~L$). Note that in YBCO the fits are under-constrained and linear regions in parameter space provide acceptable fits to the data (see maps of reduced chi-squared statistic $\chi_0^2$ in Fig.~\ref{fig:YBCO_LDep}b). However, we have assumed that the CDW order is dominated by the CuO$_2$ planes\cite{Achkar12} and imposed the constraint that $|t_{cc}/t_{\perp}| \leq 0.2$ to estimate $t_\parallel/t_{\perp}$.

This comparison identifies two key results.  Firstly, in YBCO the orbital symmetry of the CDW ordering along the $a$-axis is different than along the $b$-axis. This provides further evidence that a simple $C_4$ symmetric checkerboard order is not applicable to YBCO.\cite{Blackburn13,Blanco-Canosa13,Comin15}  Instead, this result points to CDW order with distinct unidirectional character, possibly unidirectional domains with different orbital symmetries along $a$ and $b$. Presumably the orbital symmetry difference between the $H$ and $K$ peaks ultimately results from the orthorhombic structure of YBCO, which breaks intra-unit cell $C_4$ symmetry.   However, the lack of a simple inverse relation of $|t_{\parallel}/t_{\perp}|$ between  $(\pm 0.31 ~0 ~L)$ and $(0 ~\pm 0.31 ~L)$, which would result from the intra-unit cell $C_2$ symmetry of the crystalline and electronic structure, indicates a more complex relationship between CDW orbital symmetry and structural symmetry.



Secondly, the CDW symmetry differs between 1/8 doped LBCO and YBCO.  The more asymmetric $t_\parallel/t_{\perp}$ in YBCO may result from YBCO having a larger ratio of $d$ to $s$ (and/or $s'$) symmetry CDW order than LBCO. The azimuthal dependence of YBCO at the Cu $L$ edge with Q = (0 0.31 1.48) was recently studied in terms of parameters $\delta_{d}$, $\delta_{s'}$ and  $\delta_{s}$ (that are related to but inequivalent to $\Delta_{s,s',d}$).\cite{Comin14c} This work showed that a large $\delta_{d}$ provides the best fit to the data, providing evidence for prominent $d$ symmetry in YBCO.  The orbital asymmetry identified by our measurements may be consistent with the observation of a large $d$ symmetry component to the CDW order and hint at a different, possibly larger, $d$ symmetry contribution to CDW order along the $a$-axis than along the $b$-axis.  However, we note that the $L$ dependence of CDW order does not reveal the sign of $t_\parallel/t_{\perp}$, and thus our measurements do not provide a direct and unambiguous signature of dominant $d$ symmetry CDW order in YBCO.

Moreover, an important observation of our work is that the O and Cu sublattice measurements correspond to different $t_\parallel/t_\perp$ in LBCO. This illustrates that the same CDW can exhibit different scattering anisotropies, and sensitivity to CDW orbital symmetry, from measurements of the Cu and O sublattices. This can be understood from microscopic considerations. A CDW with mixed orbital symmetry ($s+s'+d$) acts on bonds and sites differently: $\Delta_d$ and $\Delta_{s'}$ CDW modulations map directly to the bonds (O atoms) while $\Delta_s$ modulations map directly to the sites (Cu atoms). It is thus expected that Cu $L$ edge resonant scattering should be primarily sensitive to $\Delta_s$. However, since $d$ and $s'$ modulations can shift the Cu core electron energy levels,\cite{Achkar13,Efetov13} or otherwise alter the electronic orbitals involved in the $2d\to3d$ transition, they may impart their signature on the Cu sublattice scattering. At present, the relative influence at the Cu $L$ edge of site ($\Delta_s$) and bond ($\Delta_d$, $\Delta_{s'}$) centered contributions to the CDW is not known.  Consequently the relationship between $\delta_{d}/\delta_{s}$, as deduced by Comin {\it et al.} from Cu $L$ edge measurements in YBCO,\cite{Comin14c} and $\Delta_{d}/\Delta_{s}$ warrants further investigation.  

\begin{figure}
\centering
\resizebox{\columnwidth}{!}{\includegraphics{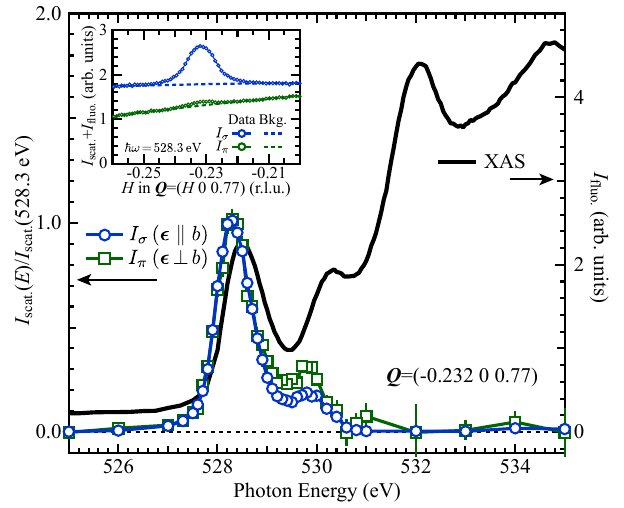}}
\caption{{\bf Photon energy dependence of scattering from O$_\parallel$ and O$_\perp$ sublattices.} RSXS scattering intensity of CDW order in LBCO with $\sigma$ (blue) and $\pi$ (green) incident photon polarization, corresponding to the O$_\perp$ and O$_\parallel$ sublattices, respectively. The scattering is scaled to 1 at 528.3 eV and compared to x-ray absorption (black) measured by total fluorescence yield with $\bm{\epsilon} \parallel \bm{a}$. Inset:  Measured intensity as a function of $H$ showing the scattering intensity along with the x-ray fluorescence background $I_\text{fluo.}$. The background is fit and subtracted to determine $I_\text{scat}$.  Error bars in the main panel are the standard deviation in the peak amplitude from Lorentzian fits to $I_\text{scat}(H)$.}
\label{fig:LBCO_EDep}
\end{figure}

The photon energy dependent sensitivity to CDW orbital symmetry observed by comparing the O $K$ and Cu $L$ edges can be examined in greater detail at the individual edges.  Our prior work showed that there is no difference in photon energy dependence for Cu $L$ edge scattering for $\sigma$ and $\pi$ polarizations in YBCO.\cite{Achkar12}  At the O $K$ edge, while the energy dependence has previously been analyzed on either the O$_{\perp}$ or the O$_{\parallel}$ sublattice,\cite{Abbamonte05,Fink09,Achkar13,Wu12,Benjamin13} here we present CDW scattering from both sublattices in LBCO. As shown in Fig.~\ref{fig:LBCO_EDep}, the main features are a primary peak at 528.3 eV and a secondary peak at 529.8 eV. Notably, $I_\pi$ (probing O$_{\parallel}$) and $I_\sigma$ (probing O$_{\perp}$) exhibit very similar energy dependence below 529 eV. This energy range corresponds to the electronic states closest to $E_\text{F}$, associated with mobile charge carriers (holes). For these states, it appears that a single symmetry configuration characterizes the CDW order.

At photon energies above 529.5 eV, the scaled scattering intensity is slightly higher for $I_\pi$ than for $I_\sigma$. Judging by the $I_\pi/I_\sigma$ ratio here, it appears that the scattering at this energy may be more isotropic ($t_{\parallel} / t_{\perp} \sim 1$) than at 528.3 eV ($t_{\parallel} / t_{\perp} \sim 0.6$).
This observation indicates that the electronic states contributing to the CDW scattering intensity at the O $K$ edge are modulated in space with intra-unit-cell orbital symmetries that depend on energy. Specifically, the electronic states at 529.8 eV appear to have much less than the 24\% $d$ symmetry identified for the states closest to $E_\text{F}$. This distinction is perhaps associated with the different character of the unoccupied states involved in the resonant scattering process; while the doped-hole states near $528.3\text{ eV}$ have dominant O $p_{x,y}$ character, those at 529.8 eV are most closely associated with the upper Hubbard band and have primarily Cu $3d$ character, being observed at the O $K$ edge due to hybridization between O $2p$ and Cu $3d$ states. These differences are not only important to the x-ray absorption of the pre-edge structure, they also appear to influence the measured symmetry the CDW order. Moreover, given the strong Cu hybridization at 529.8 eV, the nearly isotropic scattering may be related to the observation of $t_{\parallel} \approx t_{\perp}$ at the Cu $L$ edge.   

The observation of dominant $s'$ symmetry CDW order in LBCO should be contrasted with existing evidence for dominant $d$ symmetry CDW in Bi-2212 and Na-CCOC,\cite{Fujita14} and YBCO.\cite{Comin14c} If we assume both STM and RSXS have similar sensitivity to the CDW orbital symmetry, our measurements highlight CDW orbital symmetry as an additional property of CDW order distinguishing the La-based cuprates from other cuprate families. This difference may be related to the spin ordering properties of these different families - only in the La-based cuprates does static SDW order develop with an incommensurability that is clearly related to the CDW order ($\delta_\text{charge} = 2\delta_\text{spin}$).\cite{Tranquada95, Hucker11}. The related $\delta_\text{charge}$ and $\delta_\text{spin}$ in La-based cuprates is generally associated with a correlation between the local charge and the local anti-ferromagnetic (AF) ordering.\cite{Tranquada95}  For example, in a valence bond-solid description of stripes, bond-order (CDW) and SDW order can be related, as shown in Fig.~\ref{fig:ScatGeom_Models}b, where spins align ferromagnetically when the bond amplitude is sufficiently high (AF otherwise).\cite{Sachdev03,Vojta09}

We now consider how this local charge/AF correlation is influenced by different CDW orbital symmetries.  For an $s'$ symmetry CDW, the local charge/AF order for bonds parallel and perpendicular to $\bm{Q}$ are modulated in-phase. This would naturally accommodate static SDW order consistent with observations in La-based cuprates (a commensurate, bond-centered $s'$-CDW+SDW order of this type is depicted in Fig.~\ref{fig:ScatGeom_Models}b).\cite{Tranquada95,Christensen07} In contrast, $d$ symmetry bond order would have local charge/AF order for bonds parallel and perpendicular to $\bm{Q}$ modulated out-of-phase with each other. Accordingly, a $d$ symmetry CDW may prohibit static SDW order that is commensurate with the CDW, or conversely static SDW order may prohibit $d$ symmetry CDW.  This notion of a relation between CDW symmetry and static SDW order is consistent with existing theories. A recent study of CDW order in a three-orbital model that develops both spin and charge stripe order,\cite{Fischer14} reports a predominantly $s'$ symmetry CDW with a $d/s'$ proportion similar in magnitude to our findings in LBCO. In contrast, theories that have reported a $d$ symmetry CDW order have yet not exhibited both static SDW and CDW orders,\cite{Metlitski10,Sachdev13,Efetov13,Atkinson14,Yamakawa14} as found in La-based cuprates. 

We also speculate on the role of CDW symmetry in the competition between CDW order and superconductivity, for which various pictures have recently been proposed.\cite{Chowdhury14,Lee14,Meier14,Pepin14,Wang14} We note that $T_\text{c}$ is more strongly suppressed around $p=1/8$ in the LBCO than in other cuprate families. It is known that $d$-symmetry order parameters couple primarily to the anti-nodal quasiparticles, leaving nodal quasiparticles unaffected.\cite{Vojta08}  The greater competition with superconductivity in LBCO may ultimately be rooted in the symmetry of the CDW order, which may have a greater pair-breaking effect for $s'$ symmetry than for $d$ symmetry, particularly in the nodal region.

It remains an open question whether CDW order in the different cuprate families can be described within a single microscopic theory.  To this end, it will be important going forward to identify what are the key aspects of microscopic models that may differ between material families and affect the CDW orbital symmetry.

\section{\uppercase{Methods}}
RSXS and x-ray absorption spectroscopy (XAS) measurements at the Cu $L$ and O $K$ edges were performed using the in-vacuum four-circle diffractometer\cite{Hawthorn11a} at the Canadian Light Source's REIXS beam line on single-crystal samples of YBa$_2$Cu$_3$O$_{6.67}$ ($T_c = 64.5$ K, $p$ = 0.116) and La$_{1.875}$Ba$_{0.125}$CuO$_4$ ($T_c \approx 4$ K). The LBCO crystals are the same ones as reported on in a prior work.\cite{Grafe10} Reciprocal lattice units (r.l.u.) were defined using the lattice constants $a\!=\!3.84\ \text{\AA}$, $b\!=\!3.88\ \text{\AA}$, $c\!=\!11.74\ \text{\AA}$ for YBCO and $a\!=\!b\!=\!3.787\ \text{\AA}$, $c\!=\!13.24\ \text{\AA}$  for LBCO. The YBCO sample was polished and two LBCO samples were cleaved in air prior to measurement. The crystallographic orientation of the samples was verified in the diffractometer using appropriate structural Bragg peaks at $\sim$2 keV. LBCO samples were mounted separately to flat and wedge-shaped sample plugs. The first sample (S1) exhibited CDW peaks at $\bm{Q}=(\pm0.232\ 0\ L)$, whereas the second sample (S2) had them at $\bm{Q}=(\pm0.237\ 0\ L)$, indicating a slight difference in doping (incommensurability). 

For the O $K$ (Cu $L$) edge azimuthal rotation experiment on LBCO, the wedge angle $\theta_w$ was $53^\circ$ ($34^\circ$) (see Fig.~\ref{fig:ScatGeom_Models}a). Note, these wedge angles ensure that rotations of $\phi$ maintain $\bm{Q}$ constant for the same instrumental geometries $(\theta ,\omega)$ and are determined according to $\theta_w=\arctan(H/L \times c/a)$ for $(H\ 0\ L)$  scattering. Due to a limited motion range on the $\phi$ motor, manual rotations of $\phi$ were performed with an in-vacuum screwdriver. Photographs through a levelled telescope were used to measure $\phi$, yielding a precision of $\sim1^\circ$, and referenced to $\phi=0^\circ$ as set by the crystallographic orientation. The measurements on LBCO were performed at $T=22$ K whereas those on YBCO were at 60 K and 160 K. The vacuum chamber pressure for all measurements was $P < 1\times 10^{-9}$ Torr. X-ray absorption was measured by total fluorescence yield. Additional details about background subtractions, peak fitting, model calculations and parameter estimation are contained within the Supplementary Information. 

Correspondence and requests for materials should be addressed to D.G.H, email: dhawthor@uwaterloo.ca.

\acknowledgments{{\bf Acknowledgments}: The authors acknowledge insightful discussions with S. Sachdev, J.C. Davis, W. A. Atkinson, G. A. Sawatzky, R. Comin and A. Damascelli. This work was supported by the Canada Foundation for Innovation (CFI), the Canadian Institute for Advanced Research and the Natural Sciences and Engineering Research Council of Canada (NSERC). Research described in this paper was performed at the Canadian Light Source, which is funded by the CFI, the NSERC, the National Research Council Canada, the Canadian Institutes of Health Research, the Government of Saskatchewan, Western Economic Diversification Canada, and the University of Saskatchewan. J.G. and M.Z. were supported by the German Science Foundation (DFG) through the Emmy-Noether (GE1647/2-1) and the D-A-CH program (GE 1647/3-1).  The work at Brookhaven National Labs was supported by the Office of Basic Energy Sciences, Division of Materials Science and Engineering, U.S. Department of Energy, under Contract No. DE-AC02-98CH10886.Ó}

{\bf Author contributions}: A.J.A. and D.G.H. analyzed the data and wrote the manuscript. D.G.H., A.J.A., J.G., and M.H. conceived of the experiments that A.J.A., F.H., R.S., C.M., M.Z., and D.G.H. performed.  M.H., J.G., and G.D.G. provided the LBCO crystals and R.L., D.A.B., and W.N.H. provided the YBCO crystal.   

\bibliographystyle{nphys_mod}
%





\renewcommand{\thefigure}{S\arabic{figure}}
\renewcommand{\theequation}{S\arabic{equation}}
\renewcommand{\thetable}{S\Roman{table}}

\onecolumngrid
\appendix



\widetext
\setcounter{equation}{0}
\setcounter{figure}{0}
\setcounter{table}{0}
\setcounter{page}{1}
\makeatletter

\renewcommand{\thefigure}{S\arabic{figure}}
\renewcommand{\theequation}{S\arabic{equation}}
\renewcommand{\thetable}{S\Roman{table}}

\renewcommand{\bibnumfmt}[1]{[S#1]}
\renewcommand{\citenumfont}[1]{S#1}

\titlespacing\section{0pt}{3.5ex plus 1ex minus .2ex}{2.3ex plus .2ex}
\titlespacing\subsection{0pt}{3.25ex plus 1ex minus .2ex}{1.5ex plus .2ex}
\titlespacing\subsubsection{0pt}{3.25ex plus 1ex minus .2ex}{1.5ex plus .2ex}

\clearpage
\begin{center}
\textbf{\large Supplementary Information for: \\ Orbital symmetry of charge density wave order in La$_{1.875}$Ba$_{0.125}$CuO$_4$ and YBa$_2$Cu$_3$O$_{6.67}$}
\end{center}


\author{A. J. Achkar}
\affiliation{Department of Physics and Astronomy, University of Waterloo, Waterloo, N2L 3G1, Canada}
\author{F. He}
\affiliation{Canadian Light Source, Saskatoon, Saskatchewan, S7N 2V3, Canada}
\author{R. Sutarto}
\affiliation{Canadian Light Source, Saskatoon, Saskatchewan, S7N 2V3, Canada}
\author{Christopher McMahon}
\affiliation{Department of Physics and Astronomy, University of Waterloo, Waterloo, N2L 3G1, Canada}
\author{M. Zwiebler}
\affiliation{Leibniz Institute for Solid State and Materials Research IFW Dresden, Helmholtzstra{\ss}e 20, 01069 Dresden, Germany}
\author{M. H\"{u}cker}
\affiliation{Condensed Matter Physics and Materials Science Department, Brookhaven National Laboratory, Upton, NY 11973, USA}
\author{G. D. Gu}
\affiliation{Condensed Matter Physics and Materials Science Department, Brookhaven National Laboratory, Upton, NY 11973, USA}
\author{Ruixing Liang}
\affiliation{Department of Physics and Astronomy, University of British Columbia, Vancouver,V6T 1Z1, Canada}
\affiliation{Canadian Institute for Advanced Research, Toronto, Ontario M5G 1Z8, Canada}
\author{D. A. Bonn}
\affiliation{Department of Physics and Astronomy, University of British Columbia, Vancouver,V6T 1Z1, Canada}
\affiliation{Canadian Institute for Advanced Research, Toronto, Ontario M5G 1Z8, Canada}
\author{W. N. Hardy}
\affiliation{Department of Physics and Astronomy, University of British Columbia, Vancouver,V6T 1Z1, Canada}
\affiliation{Canadian Institute for Advanced Research, Toronto, Ontario M5G 1Z8, Canada}
\author{J. Geck}
\affiliation{Chemistry and Physics of Materials, Paris Lodron University Salzburg, Hellbrunner Strasse 34, 5020 Salzburg, Austria}
\author{D. G. Hawthorn}
\affiliation{Department of Physics and Astronomy, University of Waterloo, Waterloo, N2L 3G1, Canada}
\affiliation{Canadian Institute for Advanced Research, Toronto, Ontario M5G 1Z8, Canada}

\maketitle

This supplementary information is divided into four sections. First,  we describe the polarization dependent resonant scattering model used to fit the experimental data. Second, we show how the orbital symmetry, $\Delta_{s'}$ and $\Delta_d$, is mapped onto experimentally measurable quantities, $t_\parallel$ and $t_\perp$, for the O sites. Third, we present the dependence of the experimental data on $\phi$, $L$ and $\hbar \omega$. Fourth, we provide calculations that explore the parameter space of the scattering model in order to illustrate the level of confidence in our findings.

\clearpage
\section{Polarization Dependent Scattering: Model derivation and effect of absorption correction}
On an x-ray absorption edge, the resonant elastic x-ray scattering intensity is given by:
\begin{equation}
I(\bm{\epsilon}_i, \omega, \bm{Q}) \propto \left|\bm{\epsilon}_f^*\cdot T(\omega, \bm{Q}) \cdot \bm{\epsilon}_i\right|^2,
\label{eqnS1}
\end{equation}
where $\omega$ is the angular frequency, $\bm{Q}$ is the momentum transfer, $\bm{\epsilon}_i$ and $\bm{\epsilon}_f$ are the incident and scattered polarization, respectively.   $T(\omega, \bm{Q})$ is a tensorial equivalent to the structure factor, defined as 
\begin{equation}
T(\omega, \bm{Q}) = \sum_{n} F_{n}(\omega) e^{-i\bm{Q}\cdot\bm{r}_{n}} =  \left[\begin{matrix}
t_{aa} & t_{ab} & t_{ac} \\
t_{ba} & t_{bb} & t_{bc} \\
t_{ca} & t_{cb} & t_{cc} \end{matrix}\right].
\label{eqn:Ttensor}
\end{equation}
Each component represents a sum over site index $n$ (with atomic positions $\bm{r}_n$) of the atomic scattering form factor, $F_{n}(\omega) = F^{0}_{n}(\omega) + F^\text{R}_{n}(\omega)$,  where $F^{0}_n(\omega)$ and  $F^\text{R}_{n}(\omega)$ are non-resonant and resonant contributions, respectively.  The resonant part, $F^\text{R}_{n}(\omega)$, is strongly enhanced on an x-ray absorption edge and has a symmetry that captures the local symmetry of electronic structure. The atomic scattering form factor has components $f_{ij}$ with $\{i,j\}=(a,b,c)$. 

At the Cu $L$ edge, where scattering corresponds to a $2p\to 3d$ transition, a Cu atom in tetragonal CuO$_2$ planes would have
\begin{equation}
F^\text{R}_\text{Cu}(\omega) \approx  \left[\begin{matrix}
f^\text{R}_{aa}(\omega) & 0 & 0 \\
0 & f^\text{R}_{bb}(\omega) & 0 \\
0 & 0 & f^\text{R}_{cc}(\omega) \end{matrix}\right],
\label{eqn:FCu}
\end{equation}
where $f^\text{R}_{aa}(\omega) =  f^\text{R}_{bb}(\omega) \gg f^\text{R}_{cc}(\omega)$ and off-diagonal terms are negligibly small for charge scattering.  At the O $K$ edge, sites with holes in O $2p_x$ and O $2p_y$ orbitals would have different atomic scattering form factors, given by
\begin{equation}
F^\text{R}_{\text{O}2p_x} \approx  \left[\begin{matrix}
f_{aa}^\text{R}(\omega) & 0 & 0 \\
0 &0 & 0 \\
0 & 0 & 0 \end{matrix}\right], \ \ F^\text{R}_{\text{O}2p_y} \approx  \left[\begin{matrix}
0 & 0 & 0 \\
0 & f_{bb}^\text{R}(\omega)  & 0 \\
0 & 0 & 0 \end{matrix}\right].
\label{eqn:FO}
\end{equation}
In both cases, $T(\omega, \bm{Q})$ is diagonal and referenced to the crystallographic axes. It is also possible to reference the components as being either parallel or perpendicular to $\bm{Q}$, and correspondingly the orientation of the CDW. This gives the more general notation for $t_\alpha,\beta$ presented in the main text (see Eq. \ref{eqn:tpartperp}). For CDW order parallel to the $a$ and $b$ crystal axes, we would then have 
\begin{equation}
T(\omega, \bm{Q} \parallel \bm{a}^*) =  \left[\begin{matrix}
t_\parallel & 0 & 0 \\
0 & t_\perp & 0 \\
0 & 0 & t_{cc} \end{matrix}\right], \quad T(\omega, \bm{Q} \parallel \bm{b}^*) =  \left[\begin{matrix}
t_\perp & 0 & 0 \\
0 & t_\parallel & 0 \\
0 & 0 & t_{cc} \end{matrix}\right]. 
\label{eqn:Tgena}
\end{equation}

The actual measured quantities are the ratios $t_\parallel/t_\perp$ and $t_{cc}/t_\perp$ .  A key aspect of $t_\parallel/t_\perp$ is that it provides sensitivity to both the magnitude and relative phase of the modulations on the O$_\parallel$ and O$_\perp$ sublattices, which we show in the next section to relate directly to the orbital symmetry of the CDW order and distinguish $d$ and $s'$ symmetry CDW order. On the Cu sublattice, $t_\parallel/t_\perp$ is most directly associated with the orbital symmetry of the Cu $2p\to 3d$ transition.

Experimentally, $t_\parallel/t_\perp$ and $t_{cc}/t_\perp$ can be measured by rotating the sample azimuthally ($\phi$) about $\bm{Q}$ and switching the incident polarization $\bm{\epsilon}_i$ between $\sigma$ and $\pi$ (see Fig.~\ref{fig:ScatGeom_Models}a), as done in Ref. \cite{Comin14cs}. The $\phi$ rotation about $\bm{Q}$ requires mounting the crystal on a wedge with angle $\theta_w$. It is also possible to constrain $t_\parallel/t_\perp$ and $t_{cc}/t_\perp$ by varying the $L$ component of $\bm{Q}$ since the CDW scattering peak is broad in $L$ for cuprates.\cite{Abbamonte05s,Achkar14as} Our implementation of Eq. \ref{eqnS1} leaves only $t_\parallel/t_\perp$ and $t_{cc}/t_\perp$ as free parameters to be determined by fitting the data.
  
Eq.~\ref{eqnS1} references the incident and scattered photon polarization vectors relative to the crystallographic orientation of the sample. To see how variation of the experimental geometry (see Fig.~\ref{fig:ScatGeom_Models}a) affects Eq.~\ref{eqnS1}, we express the photon polarization in the laboratory reference frame (denoted by the subscript $\ell$) and Eq.~\ref{eqnS1} is rewritten as
\begin{equation}
I(\bm{\epsilon}_i,\omega, \bm{Q}) \propto \left| \bm{\epsilon}^*_{f,\ell}\cdot RT(\omega, \bm{Q}) R^\top \cdot \bm{\epsilon}_{i,\ell}\right|^2,
\label{eqn:scatgen}
\end{equation}
where $R$ is a rotation matrix that rotates the sample into the geometry necessary to satisfy the Bragg condition for a  photon energy $E=\hbar\omega$ and momentum transfer $\bm{Q}$, and can also rotate the sample azimuthally ($\phi$) about $\bm{Q}$.\cite{Comin14cs} For photon detection without polarization sensitivity, both $\bm{\epsilon}_{f,\sigma}$ and $\bm{\epsilon}_{f,\pi}$ scattering contribute to the scattering intensity, giving $I(\bm{\epsilon}_{i}) = I_{\bm{\epsilon}_{i},\sigma '} + I_{\bm{\epsilon}_{i},\pi '}$.

To calculate the scattering intensity for a given symmetry of $T(\omega, \bm{Q})$ using Eq.~\ref{eqn:scatgen}, the rotations that the sample and photon polarization undergo relative to the lab reference frame must be specified.  We incorporate a lab reference frame with $\bm{k}_i$ along the (1 0 0)$_\ell$ direction and the scattering plane being orthogonal to the (0 0 1)$_\ell$ direction, giving $\bm{\epsilon}_{i,\sigma}$ = (0 0 1)$_\ell$ and $\bm{\epsilon}_{i,\pi}$ = (0 1 0)$_\ell$.  The sample can be mounted on a wedge, defined by a wedge angle, $\theta_w$, ($\theta_w = 0$ corresponds to a flat surface, i.e. no wedge) that can be rotated azimuthally about (0 1 0)$_\ell$ by an angle $\phi$.  The azimuthally rotated wedge can then be rotated by an angle $\theta$ about the (0 0 1)$_\ell$ axis.  For $\theta_w = 0$, $\phi = 0$ and  $\theta = 0$, the samples are mounted such that the crystalline axes $c$ // (0 1 0)$_\ell$ and $a$ [$b$] along (1 0 0)$_\ell$ for investigation of ($H$ 0 $L$) [($0$ $K$ $L$)] peaks.  These rotations can be expressed as a rotation matrix $R$ acting on $T$ given by
\begin{equation}
R = R(\theta)_{0 0 1}R(\phi)_{0 1 0}R(\theta_w)_{0 0 1}R_O,
\label{eqn:Rmatrix}
\end{equation}
where, for example, $R(\theta)_{0 0 1}$ rotates the sample by an angle $\theta$ about the (0 0 1)$_\ell$ axis and $R_O$ provides an initial orientation of the sample to give $c \parallel (0~1~0)_\ell$ and $a$ or $b$ along (1 0 0)$_\ell$.   The scattered photon polarization,  $\bm{\epsilon}_f$,  is determined by $\bm{\epsilon}_i$ and $RTR^\top$ and can be expressed in terms of $\bm{\epsilon}_{f,\sigma}$ and $\bm{\epsilon}_{f,\pi}$ as 
\begin{equation}
 \bm{\epsilon}_f = \frac{(\bm{\epsilon}_{f,\sigma}^* \cdot RTR^\top \cdot \bm{\epsilon}_{i})\bm{ \epsilon}_{f,\sigma} +  (\bm{\epsilon}_{f,\pi}^* \cdot RTR^\top \cdot \bm{\epsilon}_{i})\bm{\epsilon}_{f,\pi}}{|(\bm{\epsilon}_{f,\sigma}^* \cdot RTR^\top \cdot \bm{\epsilon}_{i})\bm{ \epsilon}_{f,\sigma} +  (\bm{\epsilon}_{f,\pi}^* \cdot RTR^\top \cdot \bm{\epsilon}_{i})\bm{\epsilon}_{f,\pi}|}.
 \label{eqn:epsilonf}
\end{equation}
where $\bm{\epsilon}_{f,\sigma} = (0 ~0 ~1)_\ell$, $\bm{\epsilon}_{f,\pi} = (-\sin \Omega~\cos \Omega  ~0)_\ell$  and $\Omega$ is the detector angle relative to the incident beam.  The scattering intensity can be determined without knowledge of $\bm{\epsilon}_f$ by 
\begin{equation}
I(\bm{\epsilon}_{i}, \omega, \bm{Q}) \propto \left| \bm{\epsilon}^*_{f,\sigma} \cdot RT(\omega, \bm{Q}) R^\top\cdot \bm{\epsilon}_{i}\right|^2+\left| \bm{\epsilon}^*_{f,\pi}\cdot RT(\omega, \bm{Q}) R^\top\cdot \bm{\epsilon}_{i}\right|^2.
\label{eqn:scatsumepsilons}
\end{equation}

When comparing experimental results to model calculations, it is also important to account for the polarization dependent x-ray absorption cross-section, $\mu(\omega,\bm{\epsilon})$, of the incident and scattered photons.  This is a minor correction in more electronically isotropic materials, but can have a significant impact in the cuprates, where the absorption coefficient along the $a$, $b$ and $c$ axes varies considerably.  To account for geometry dependent attenuation of the incident and scattered x-rays, Eq.~\ref{eqn:scatsumepsilons} must be corrected according to
\begin{equation}
I_\text{abs}(\bm{\epsilon}_{i}, \omega, \bm{Q})\propto\frac{I(\bm{\epsilon}_{i}, \omega, \bm{Q})}{\mu_i +\mu_f \frac{\sin \alpha}{\sin \beta}},
\label{eqn:scatcorrabs}
\end{equation}
where $\alpha$ and $\beta$ are the angles of the incident and scattered beam relative to the sample surface in the scattering plane and $\mu_i$ and $\mu_f$ are the linear absorption coefficients of the incident and scattered photons respectively.  These are given by 
\begin{eqnarray}
 \mu_i &\propto& \text{Im}\!\left(\bm{\epsilon}_i^* R \overline{F} R^\top \bm{\epsilon}_i\right)  \label{eqn:mui}  \\
 \mu_f &\propto& \text{Im}\!\left(\bm{\epsilon}_f^* R  \overline{F} R^\top \bm{\epsilon}_f \right) \label{eqn:muf}
 \end{eqnarray}
where $\overline{F}$ is the scattering tensor averaged over all atomic sites (O, Ba, Cu, ...) and includes both resonant and non-resonant contributions.  Mirroring the polarization dependent x-ray absorption in the cuprates \cite{Chen92s,Hawthorn11bs,Achkar13s}, $\text{Im}(\overline{f_{aa}}) \simeq \text{Im}(\overline{f_{bb}})$ and we estimate $\text{Im}(\overline{f_{cc}})/\text{Im}(\overline{f_{aa}}) \simeq$ 0.74 at the O $K$ edge (528.3 eV) in LBCO, 0.45 at the Cu $L$ edge (931.4 eV) in LBCO and 0.66 at the Cu $L$ edge (931.4 eV) in YBCO. \cite{Hawthorn11bs,Achkar13s} For these estimates in LBCO, we assume that the absorption coefficient is similar to that of LNSCO, where polarization dependent XAS is available.  

\begin{figure}
\centering
\resizebox{\columnwidth}{!}{\includegraphics{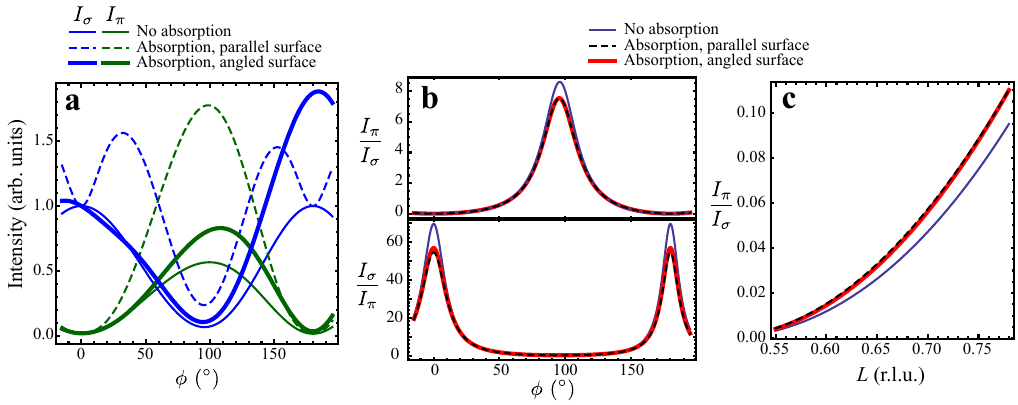}}
\caption{Model calculations of the $\phi$ and $L$ dependence of the scattering intensity in LBCO at the O $K$ edge ($t_{\parallel}/t_{\perp}=0.612$, $t_{cc}/t_\perp=0.034$). {\bf a.} $I_\pi$ and $I_\sigma$ vs. $\phi$ calculated by Eq.~\ref{eqn:scatsumepsilons} (no absorption, thin line) and Eq.~\ref{eqn:scatcorrabs} (absorption) for a parallel surface (dashed line) and the angled surface of the measured sample (thick line). The ratios of $I_\pi$ and $I_\sigma$ vs. $\phi$ ({\bf b}) and $L$ ({\bf c}).}
\label{fig:Supp_ModelAbsorption}
\end{figure}

The effect of the absorption correction on the $\phi$ and $L$ dependence of the model are illustrated in Fig.~\ref{fig:Supp_ModelAbsorption} for LBCO at the O $K$ edge. The calculation without any absorption correction (Eq.~\ref{eqn:scatsumepsilons}) is shown as thin solid lines. The effect of including the absorption correction (Eq.~\ref{eqn:scatcorrabs}) is shown with dashed lines. For the usual case where the sample surface is parallel to the wedge surface, the dashed line would be the ideal calculation. However, cleaving the sample LBCO S1 yielded a surface that was not parallel to the sample holder (this can also occur for intentionally miscut surfaces), affecting the angles $\alpha$ and $\beta$ that enter into Eq.~\ref{eqn:scatcorrabs}. The actual surface orientation was estimated with an optical microscope and included in the model calculations, shown as thick lines. The $\phi$ dependence of $I_\pi$ and $I_\sigma$ (Fig.~\ref{fig:Supp_ModelAbsorption}a) changes significantly when the absorption correction and the orientation of the surface are incorporated. Examination of Fig.~\ref{fig:Supp_ModelAbsorption}a illustrates that modelling the azimuthal dependence of $I_\pi$ and $I_\sigma$ directly (as shown in Fig.~\ref{fig:LBCO_PolDep}a in the main text) requires a full characterization of the experimental geometry and the effect of absorption on the incident and scattered photons. 

When evaluating the ratio of $I_\pi$ and $I_\sigma$, however, the important differences seen in Fig.~\ref{fig:Supp_ModelAbsorption}a are largely factored out, as shown in Figs.~\ref{fig:Supp_ModelAbsorption}b and \ref{fig:Supp_ModelAbsorption}c. The effect of the absorption correction is modest, on the order of $\sim$ 5\%-15\% near $\phi=0^\circ ,97^\circ , 180^\circ$, but can be important in accurately determining the parameters $t_\parallel/t_\perp$ and $t_{cc}/t_\perp$. The effect of the angled surface is seen to be less important in calculating the ratios of $I_\pi$ and $I_\sigma$ (Figs.~\ref{fig:Supp_ModelAbsorption}b and \ref{fig:Supp_ModelAbsorption}c). This highlights that our estimate of the sample surface orientation (due to the cleave) is not important in the determination of $t_\parallel/t_\perp$ and $t_{cc}/t_\perp$ from fits to the ratio $I_\pi/I_\sigma$.

\clearpage
\section{Orbital symmetry of CDW order on oxygen sites}

Following Ref.~\cite{Sachdev13s}, non $s$-wave symmetry CDW order in a single-band model of the CuO$_2$ planes can be parameterized by $\Delta_{ij}$, which characterizes the charge modulation (or some other quantity related to charge order such as an energy shift)\cite{Achkar12s,Achkar13s} on bonds connecting nearest neighbour Cu sites $i$ and $j$:
\begin{equation}
\Delta_{ij} = \sum_{\bm{Q^*}} \left[ \frac{1}{V} \sum_{\bm{k}} e^{i \bm{k} \cdot \left( \bm{r}_i - \bm{r}_j \right)} \Delta_{\bm{Q^*}}(\bm{k}) \right] e^{i\bm{Q^*}\cdot \left( \bm{r}_i + \bm{r}_j \right)/2 },
\label{eqn:deltaij}
\end{equation}
where $V$ is volume and $\bm{Q^*}$ are the wave vectors of the CDW order: $\bm{Q^*}$ = ($\pm Q_m$, 0, 0) or (0, $\pm Q_m$, 0) for 1D stripes and $\bm{Q^*}$ = ($\pm Q_m$, 0, 0) and (0, $\pm Q_m$, 0) for checkerboard order.  In this prescription, $\bm{r}_i$ indexes the Cu sites so that $\Delta_{ii}$ corresponds to Cu sites and $\Delta_{ij}$ with $i \neq j$ corresponds to bonds between Cu sites (i.e. O sites).  Including the $s$, $s'$ and $d$ symmetry terms, 
\begin{equation}
\Delta_{\bm{Q^*}}(\bm{k})  = \left\{
\begin{array}{l l}
\Delta_s +\Delta_{s'} (\cos k_xa +\cos k_yb) + \Delta_d (\cos k_xa -\cos k_yb) & \quad \bm{Q^*} = (\pm Q_m, 0, 0)	\\
\Delta_s +\Delta_{s'} (\cos k_xa +\cos k_yb) - \Delta_d (\cos k_xa -\cos k_yb) & \quad \bm{Q^*} = (0, \pm Q_m, 0)
\end{array}
\right.
\label{eqn:deltaQstar}
\end{equation}
giving rise to spatial modulations of $\Delta_{ij}$ given in Fig.~\ref{fig:ScatGeom_Models}b.  

This model identifies two distinct sublattices for the bonds, $ij$:  one set of bonds, $A$, having $\bm{r}_j = \bm{r}_i \pm a \hat{x}$ and the other set, $B$, having $\bm{r}_j = \bm{r}_i \pm b \hat{y}$.  For these two sublattices, a stripe state with $\bm{Q^*}$ = ($\pm Q_m$, 0, 0) has
\begin{equation}
\begin{array}{l}
\Delta_{iA} = (\Delta_{s'} + \Delta_d)\cos[Q_m(r_{ix} + a/2)]\\
\Delta_{iB} = (\Delta_{s'} - \Delta_d)\cos[Q_mr_{ix}],
\end{array}
\label{eqn:deltaiABstripe}
\end{equation}
whereas a checkerboard with $\bm{Q^*}$ = ($\pm Q_m$, 0, 0) and $\bm{Q^*}$ = (0, $\pm Q_m$, 0) has
\begin{equation}
\begin{array}{l}
\Delta_{iA} = (\Delta_{s'} + \Delta_d)\cos[Q_m(r_{ix} + a/2)] + (\Delta_{s'} - \Delta_d)\cos[Q_mr_{iy}]\\
\Delta_{iB} = (\Delta_{s'} - \Delta_d)\cos[Q_mr_{ix}] + (\Delta_{s'} + \Delta_d)\cos[Q_m(r_{iy}+ b/2)] .
\end{array}
\label{eqn:deltaiABcheckerboard}
\end{equation}

Note, although \ref{eqn:deltaij} was developed in Ref.~\onlinecite{Sachdev13s} for a microscopic model based on a single-band theory, this parameterization of the CDW symmetry is sufficiently general can and can been used to describe the intra-unit cell orbital symmetry of CDW order in a 3-band model \cite{Fischer14s,Thomson14s} and CDW order having different microscopic origins.

Translating this model into bond order on the O atoms, one can associate $\Delta_{iA}$ and $\Delta_{iB}$ with O atoms having holes in 2$p_x$  and 2$p_y$ orbitals, respectively. 
Next, we assume that $\Delta_{iA(B)}$ are proportional to modulations in the atomic scattering form factor $\Delta F_{O,iA(B)}(\omega)$ giving
\begin{equation}
\Delta F_{O,iA}(\omega) = C(\omega) \left[\begin{matrix}
\Delta_{iA} & 0 & 0 \\
0 & 0 & 0 \\
0 & 0 & 0 \end{matrix}\right]\text{\ \ \ and \ \ \ } \Delta F_{O,iB}(\omega) = C(\omega) \left[\begin{matrix}
0 & 0 & 0\\
0 & \Delta_{iB}  & 0 \\
0 & 0 & 0 \end{matrix}\right],
\label{eqn:deltaiFOiAiB}
\end{equation} 
where $C(\omega)$ is a proportionality constant common to $\Delta F_{O,iA}(\omega)$ and $\Delta F_{O,iB}(\omega)$. For scattering at the O $K$ edge in LBCO, the assumption that the energy dependence factorizes from the modulation amplitude and is common to the $A$ and $B$ sublattices seems justified given the approximate proportionality between the $I_\pi$ and $I_\sigma$ scattering (see Fig.~\ref{fig:LBCO_EDep}).  Then, for a measured $\bm{Q}$ = ($Q_m$ 0 $L$), 
\begin{eqnarray}
 t_{\parallel} &=& t_{aa} = \frac{C(\omega)}{N}\sum_i (\Delta_{s'} + \Delta_d)\cos[Q_m(r_{ix}+a/2)]e^{iQ_m (r_{ix}+a/2)} \text{\ and} \label{eqn:scatgen4} \\
 t_{\perp} &=& t_{bb} = \frac{C(\omega)}{N}\sum_i (\Delta_{s'} - \Delta_d)\cos[Q_mr_{ix}]e^{iQ_mr_{ix}}, \label{eqn:tperp}
\end{eqnarray}
where $r_{ix}$ is the position of the Cu sites.  Combining Eqs. \ref{eqn:scatgen4} and \ref{eqn:tperp} gives 
\begin{equation}
\frac{t_{\parallel}}{t_{\perp}} = \frac{\Delta_{s'} + \Delta_d}{\Delta_{s'} - \Delta_d},
 \label{eqn:tparraoverperp}
\end{equation}
which is written in terms of $\Delta_{d}/\Delta_{s'}$ in Eq. \ref{eqn5}, parametrizing how the experimental observables, $t_{\parallel}$ and $t_{\perp}$, can be mapped to the symmetry of the CDW order for O sites in the CuO$_2$ plane.

\section{Influence of CDW orbital symmetry and $c$ axis scattering contribution on model}

\begin{figure}[h]
\centering
\resizebox{0.75\columnwidth}{!}{\includegraphics{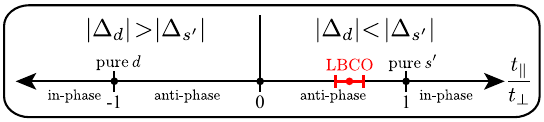}}
\caption{Mapping between CDW orbital symmetry and $t_\parallel/t_\perp$ for O sublattice scattering based on Eq. \ref{eqn5}. For $t_\parallel/t_\perp<0$ the $\Delta_d$ symmetry term dominates and for $t_\parallel/t_\perp>0$ the $\Delta_{s'}$ symmetry term dominates. The pure symmetry cases ($|t_\parallel/t_\perp|=1$) separate regions of in-phase and anti-phase $d+s'$ symmetry. The experimental result for LBCO is indicated in red. }
\label{fig:Supp_Mapping}
\end{figure}

\begin{figure}
\centering
\resizebox{\columnwidth}{!}{\includegraphics{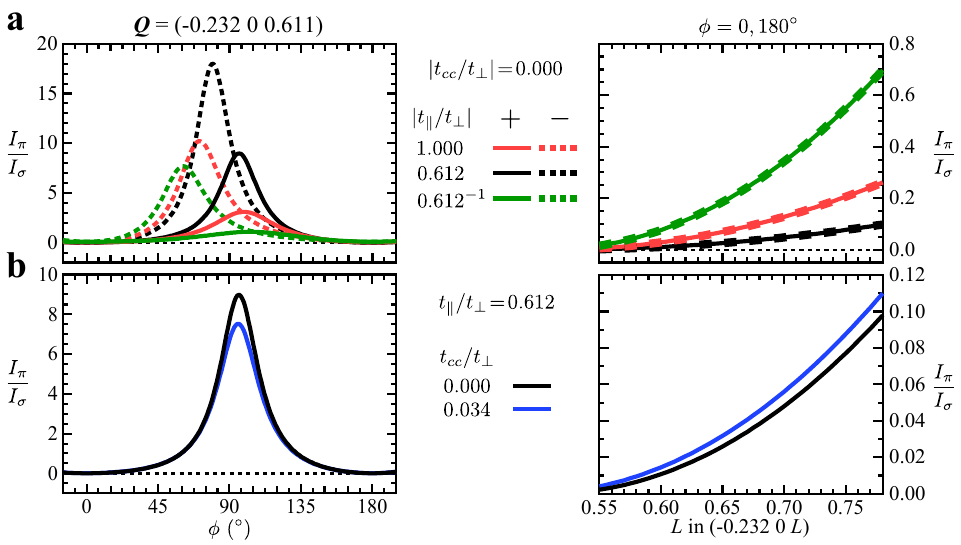}}
\caption{Model calculations for different cases of orbital symmetry in LBCO at the O $K$ edge. {\bf a.} Calculated $\phi$ (left panels) and $L$ (right panels) dependence of $I_\pi/I_\sigma$ showing sensitivity to the sign and magnitude of $t_{\parallel}/t_{\perp}$. {\bf b.} Calculated $\phi$ and $L$ dependence of $I_\pi/I_\sigma$ demonstrating sensitivity to the $c$ axis contribution $t_{cc}/t_\perp$ of $T$. The blue curve is the best fit to the data.}
\label{fig:Supp_ModelComparison}
\end{figure}

The sensitivity of the model to $\phi$ and the scattering geometry is most readily understood through calculations for representative sets of parameters corresponding to different CDW orbital symmetries. In Fig.~\ref{fig:Supp_ModelComparison} we show the $\phi$ and $L$ dependence of $I_\pi/I_\sigma$ calculated according to Eq.~\ref{eqn:scatcorrabs} for values of $t_\parallel/t_\perp=\pm1,\pm0.612,\pm0.612^{-1}$. We can interpret these cases using Eq. \ref{eqn5}. A visual guide to interpreting Eq. \ref{eqn5} is given in Fig. \ref{fig:Supp_Mapping}. 

The case $t_\parallel/t_\perp=+1$ ($-1$) corresponds to pure $s'$ ($d$) CDW symmetry (red lines in Fig.~\ref{fig:Supp_ModelComparison}). We see that $d$ symmetry (red, dashed) would give rise to $\sim3.3\times$ larger maximum in $I_\pi/I_\sigma$ than $s'$ symmetry (red, solid) and the maximum in $\phi$ would be located at $\phi\approx71^\circ$ rather than $\phi\approx99^\circ$. The case $t_\parallel/t_\perp=+0.612$ ($-0.612$) corresponds to $\Delta_d / \Delta_{s'} = -0.241$ ($-0.241^{-1}$). This is an \textit{anti-phase}, mixed $d$ and $s'$ state, with mostly $s'$ ($d$) symmetry, which would appear experimentally as a maximum in $I_\pi/I_\sigma$ near $\phi\approx79^\circ$ ($\phi\approx96^\circ$). The mostly $d$ symmetry (black, dashed) case would give rise to a $\sim2\times$ larger maximum in $I_\pi/I_\sigma$ than mostly $s'$ symmetry (black, solid).  Similarly, $t_\parallel/t_\perp=+0.612^{-1}$ ($-0.612^{-1}$) corresponds to $\Delta_d / \Delta_{s'} = 0.241$ ($0.241^{-1}$). This is an \textit{in-phase}, mixed $d$ and $s'$ state, with mostly $s'$ ($d$) symmetry. Again, the different peak positions and maximum values of $I_\pi/I_\sigma$ would distinguish between the mostly $s'$ and $d$ cases.

\vspace{10 pt}

\clearpage

From this discussion and examination of Fig.~\ref{fig:Supp_ModelComparison}a, it becomes apparent that the $\phi$ dependence is sensitive to the sign and magnitude of $t_\parallel/t_\perp$. In contrast, the $L$ dependence is sensitive to the magnitude but not the sign of $t_\parallel/t_\perp$. Combining both types of measurements enhances the reliability of experimentally determining the sign and magnitude of $t_\parallel/t_\perp$, ultimately enabling the determination of $\Delta_d / \Delta_{s'}$. This sensitivity to the magnitude and sign of $\Delta_d / \Delta_{s'}$ affirms that polarization dependent resonant soft x-ray scattering  is a powerful tool in discerning the symmetry of CDW order in the cuprates.

In Fig.~\ref{fig:Supp_ModelComparison}b, we illustrate how a $c$ axis contribution to the scattering influences the model calculation. The blue curves are the best fit to the experimental data on LBCO at the O $K$ edge. The effect of a non-zero $t_{cc}/t_\perp$ is modest and similar in magnitude to the effect of the absorption correction seen in Figs.~\ref{fig:Supp_ModelAbsorption}b and \ref{fig:Supp_ModelAbsorption}c. This highlights the importance of including a full geometry dependent absorption correction in the model, as not doing so could lead to erroneous determinations of $t_\parallel/t_\perp$ and $t_{cc}/t_\perp$.

\clearpage
\section{Experimental data}
\begin{figure}[h]
\centering
\resizebox{\columnwidth}{!}{\includegraphics{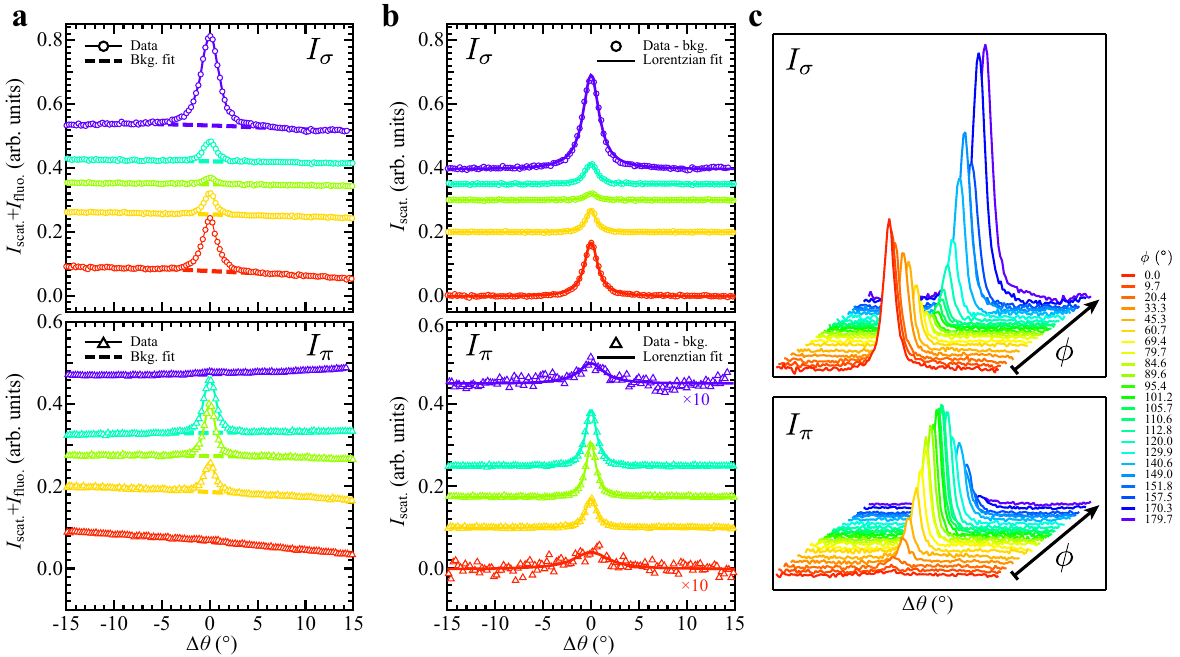}}
\caption{Scans through the CDW peak as a function of $\phi$ for LBCO at the O $K$ edge. {\bf a.} Normalized intensity of scattering and x-ray fluorescence (open symbols) for incident $\sigma$ (top panel) and $\pi$ (bottom panel) photon polarization and polynomial background fit (dashed lines) for select $\phi$ values.  {\bf b.} Background-subtracted scattering intensity $I_\sigma$ (open circles) and $I_\pi$ (open triangles), and corresponding Lorenztian fits (solid lines). In {\bf a} and {\bf b} the data is offset vertically for clarity. {\bf c.} Projected view of background-subtracted $I_\sigma$ (top) and $I_\pi$ (bottom) vs. $\phi$ for full range of measured angles.  $\phi$ values are indicated in legend on right.}
\label{fig:Supp_PhiDep_O}
\end{figure}

Here we provide the experimental data that was used to determine the $\phi$ and $L$ dependences of $I_\sigma$ and $I_\pi$ shown in the main text. Fig.~\ref{fig:Supp_PhiDep_O} shows the $\phi$ dependence of $I_\sigma$ and $I_\pi$ in LBCO at the O $K$ edge ($\bm{Q}=(-0.232~0~0.611)$, $\hbar\omega=528.3~\text{eV}$). For each scan, the detector position (at angle $\Omega$) was kept constant and the crystal was rotated about the vertical axis ($\perp$ to the scattering plane) by $\pm15^\circ$. $I_\sigma$ and $I_\pi$ were determined by first subtracting the x-ray fluorescence background ($I_\text{fluo.}$) using a polynomial fit that excluded the peak region and then fitting the resulting data ($I_\text{scat.}$) with a Lorentzian curve whose amplitude is reported in Fig.~\ref{fig:LBCO_PolDep}a in the main text. Figs.~\ref{fig:Supp_PhiDep_O}a and \ref{fig:Supp_PhiDep_O}b show this analysis procedure for a representative set of measurements. Fig.~\ref{fig:Supp_PhiDep_O}c shows a projected view of the background-subtracted $I_\sigma$ and $I_\pi$ data for the full range of $\phi$ used in fitting to the scattering model.

Here we plot this data against $\Delta\theta=\theta-\theta_0$, where $\theta_0$ is the center of the peak as identified by peak fitting. Due to a slight misalignment of the $a$ and $c$ crystal axes during sample mounting ($\sim0.6^\circ$ about wedge normal), the scattering geometry gradually shifted away from the nominal scattering geometry as $\phi$ was varied. This movement was small enough that minor adjustments to the instrument's $\chi$ angle ($<1.5^\circ$) could be used to reposition the CDW peak in the scattering plane, but it did moderately shift the apparent peak position in $\theta$ as $\phi$ was adjusted (the maximum deviation was $\sim 1^\circ$). By plotting against $\Delta\theta$, we account for these minor shifts. 
\\
\\
\\
\\

\begin{figure}[h]
\centering
\resizebox{\columnwidth}{!}{\includegraphics{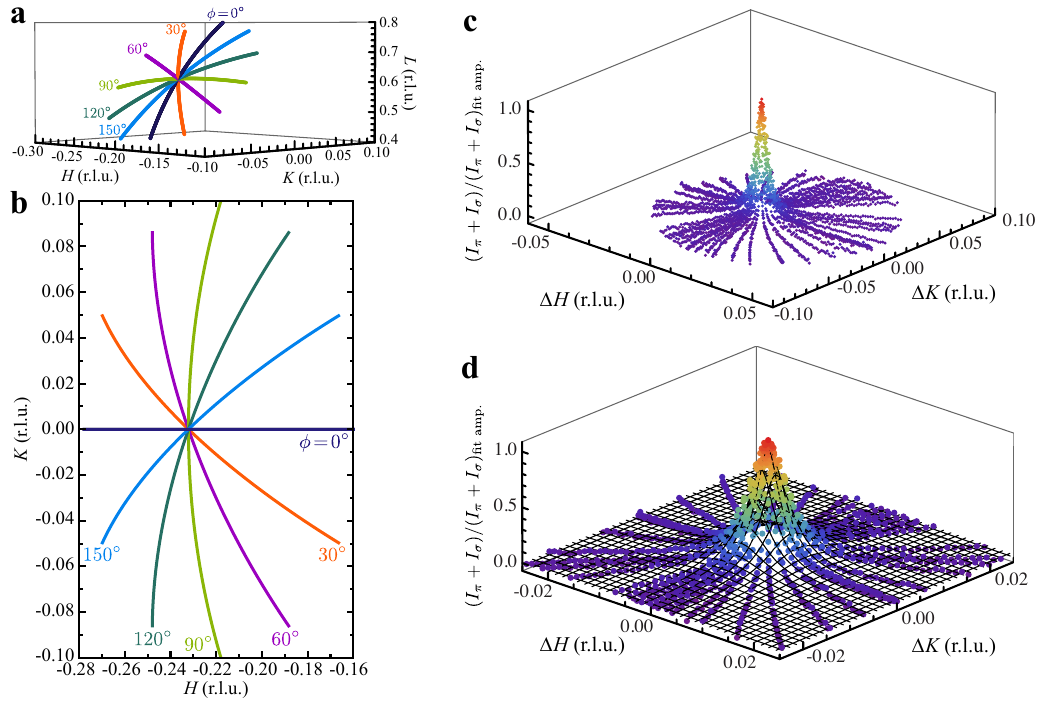}}
\caption{Cuts in ${\bm Q}$ space and $\phi$ dependence of $I_\pi+I_\sigma$ in $H$--$K$ plane. {\bf a.} Depiction of cuts in $H$, $K$, and $L$ when $\theta$ is varied by $\pm 15^\circ$ and $\Omega$ is kept fixed for indicated $\phi$ values. {\bf b.} Projection of these cuts onto the $H$--$K$ plane. {\bf c.} $I_\pi+I_\sigma$ data, normalized to the sum of their fit amplitudes, plotted in $H$--$K$ plane, and shifted to the origin. {\bf d.} A Lorentzian fit (mesh) to the data giving the reported correlation lengths in $H$ and $K$.}
\label{fig:Supp_PhiDep3D}
\end{figure}

\clearpage
The measurement scheme we have employed ($\Omega$ remains fixed while $\theta$ is rotated) has the advantage of faster data collection speed and reduced noise, but a consequence is that the $H$, $K$ and $L$ indices all vary to differing degrees that depend on $\phi$. These scans thus correspond to cuts in ${\bm Q}$ space, as illustrated in Fig.~\ref{fig:Supp_PhiDep3D}a, that depend on $\phi$. Fig.~\ref{fig:Supp_PhiDep3D}b shows a projection onto the $H$--$K$ plane of these same cuts. As can be seen, a scan at $\phi=0^\circ$ is mostly along $H$, with no $K$ and some $L$ variation, whereas one at $\phi=90^\circ$ is mostly along $K$, with some $H$ and $L$ variation. 

In Fig.~\ref{fig:Supp_PhiDep3D}c, we present the same data as in Fig.~\ref{fig:Supp_PhiDep_O} but here we have converted $\theta$ values into $H$, $K$ and $L$ indices and plotted the normalized sum of $I_\pi$ and $I_\sigma$ scattering in the $H$--$K$ plane. $H$ and $K$ were shifted by $H_0$ and $K_0$ in order to account for the shifts in $\theta$ discussed above (ie. the data is plotted against $\Delta H$ and $\Delta K$). The normalization is to the sum of the fit amplitudes of the $I_\pi$ and $I_\sigma$ data. In this form, the data can be fit to a Lorentzian function $I=A \left[(\Delta H/\gamma_H)^2+(\Delta K/\gamma_K)^2+1\right]^{-1}$, where $A$ is the amplitude and $\gamma_{(H,K)}$ is the HWHM. The $H$ and $K$ correlation lengths are related to $\gamma_{(H,K)}$ according to $\xi_{(H,K)}=(a,b)/(2\pi\gamma_{(H,K)})$. Fitting the normalized  $I_\pi+I_\sigma$ data with this functional form (see Fig.~\ref{fig:Supp_PhiDep3D}d) gives $\xi_{H}\approx 215$ \AA\ and $\xi_{K}\approx 184$ \AA. These correlation lengths and their $\sim17\%$ anisotropy appear consistent with prior hard x-ray scattering data on LBCO.\cite{Kim08s} 
\\
\\
\\
\\

\begin{figure}[h]
\centering
\resizebox{\columnwidth}{!}{\includegraphics{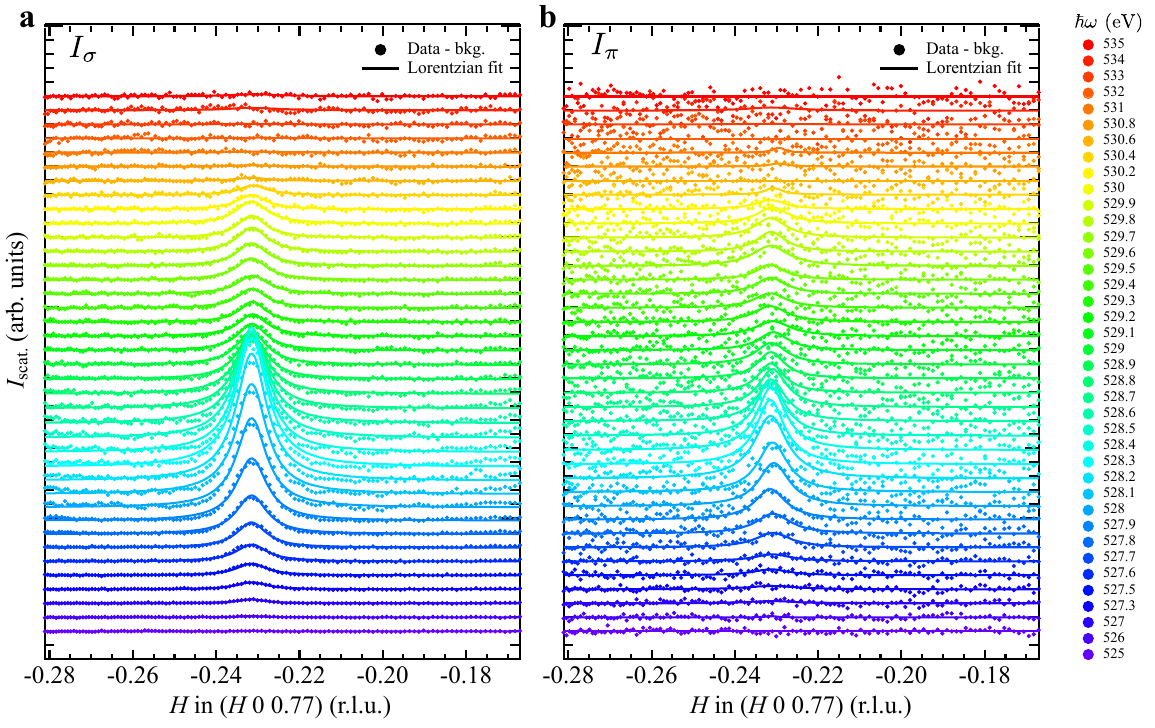}}
\caption{Energy dependence of CDW scattering peak in LBCO at the O $K$ edge for $\sigma$ ({\bf a}) and $\pi$ ({\bf b}) incident photon polarization. Solid lines are Lorenztian fits to the background-subtracted data (filled circles). Photon energy indicated in legend on the right. Data are offset for clarity.}
\label{fig:Supp_RawEnergyLBCOOK}
\end{figure}

\clearpage
Fig.~\ref{fig:Supp_RawEnergyLBCOOK} shows the photon energy dependence of $I_\sigma$ and $I_\pi$ vs. $H$ for LBCO at the O $K$ edge with $\bm{Q}=(H~0~0.77)$. Similar to the $\phi$ dependent data, the x-ray fluorescence background was first subtracted using a polynomial fit excluding the peak region (the backgrounds shown in Fig.~\ref{fig:Supp_PhiDep_O}a are representative of this procedure). Lorentzian fits were then used to determine the scattering intensities reported in Fig.~\ref{fig:LBCO_EDep}. 
\\ \\ \\

\clearpage

\begin{figure}[h]
\centering
\resizebox{\columnwidth}{!}{\includegraphics{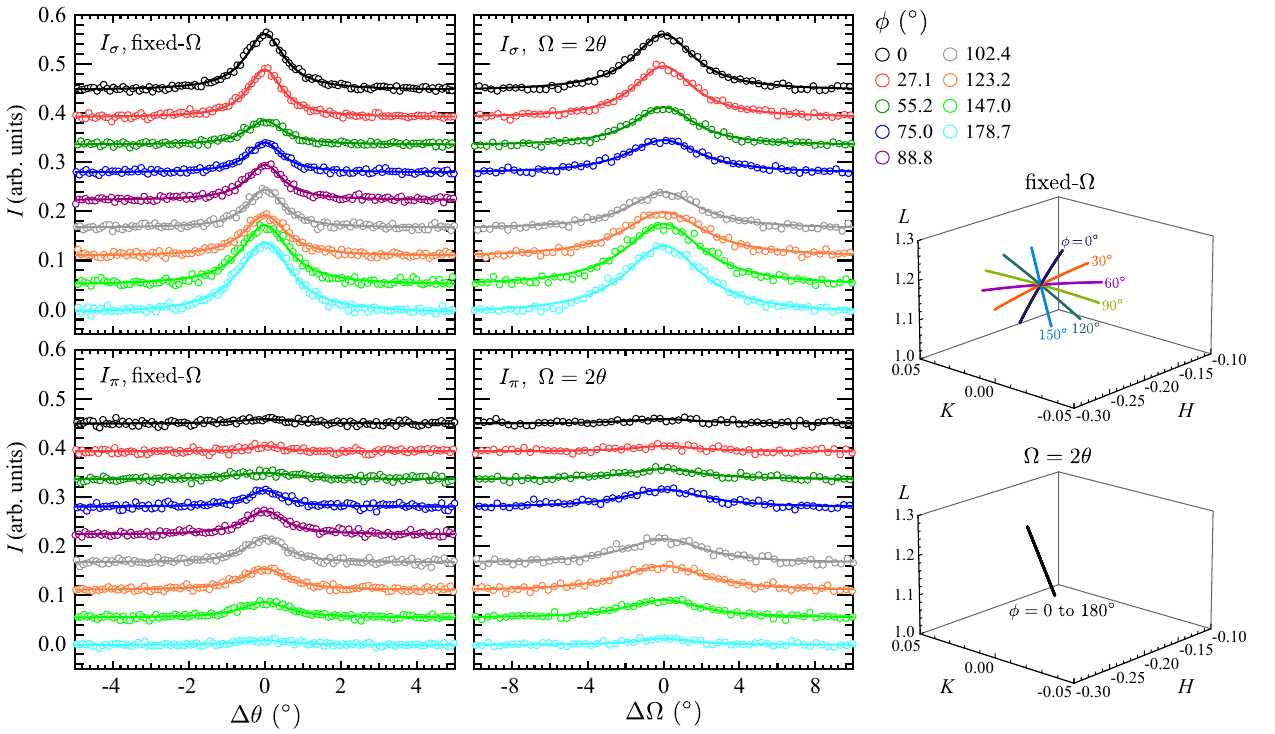}}
\caption{Scans through the CDW peak as a function of $\phi$ for LBCO at the Cu $L$ edge. The top (bottom) panels are for $I_\sigma$ ($I_\pi$) and the left (middle) panels are for fixed-$\Omega$ ($\Omega=2\theta$) scans at the $\phi$ values indicated in the top right legend. The cuts through $\bm{Q}$ space for fixed-$\Omega$ and $\Omega=2\theta$ scans are illustrated in the lower right;  }
\label{fig:Supp_PhiDep_Cu}
\end{figure}

Fig.~\ref{fig:Supp_PhiDep_Cu} shows the $\phi$ dependence of $I_\sigma$ (top panels) and $I_\pi$ (lower panels) of LBCO at the Cu $L$ edge, $\bm{Q}=(-0.236~0~1.192)$, $\hbar\omega=931.4~\text{eV}$. As was done for the O $K$ edge, the x-ray fluorescence background was subtracted using a polynomial fit excluding the peak region. Similar to the O $K$ edge $\phi$ rotation data, we plot the Cu $L$ edge data against $\Delta\theta$ to account for shifts in the scattering geometry as $\phi$ was varied (in this case, a maximum shift in $\theta$ of $\sim 5^\circ$ and $\chi$ corrections up to $\sim 4^\circ$ were used, although the mounting error was {\it smaller} for this measurement [$\sim 0.4^\circ$], there is a greater sensitivity to mounting errors at 931.4 eV as compared to 528.5 eV). In addition to scans with $\Omega$-fixed, we performed scans with $\Omega=2\theta$ and these are plotted against $\Delta\Omega$. These latter scans correspond to continually measuring the same arc in the $H$--$L$ plane whereas the $\Omega$-fixed scans correspond to arcs that vary with $\phi$, as shown in the right panels of Fig.~\ref{fig:Supp_PhiDep_Cu}. The $I_\pi$ and $I_\sigma$ values reported in Fig.~\ref{fig:LBCO_PolDep}b are the average Lorenztian fit amplitude of both types of scan. 

\begin{figure}[h]
\centering
\resizebox{0.82\columnwidth}{!}{\includegraphics{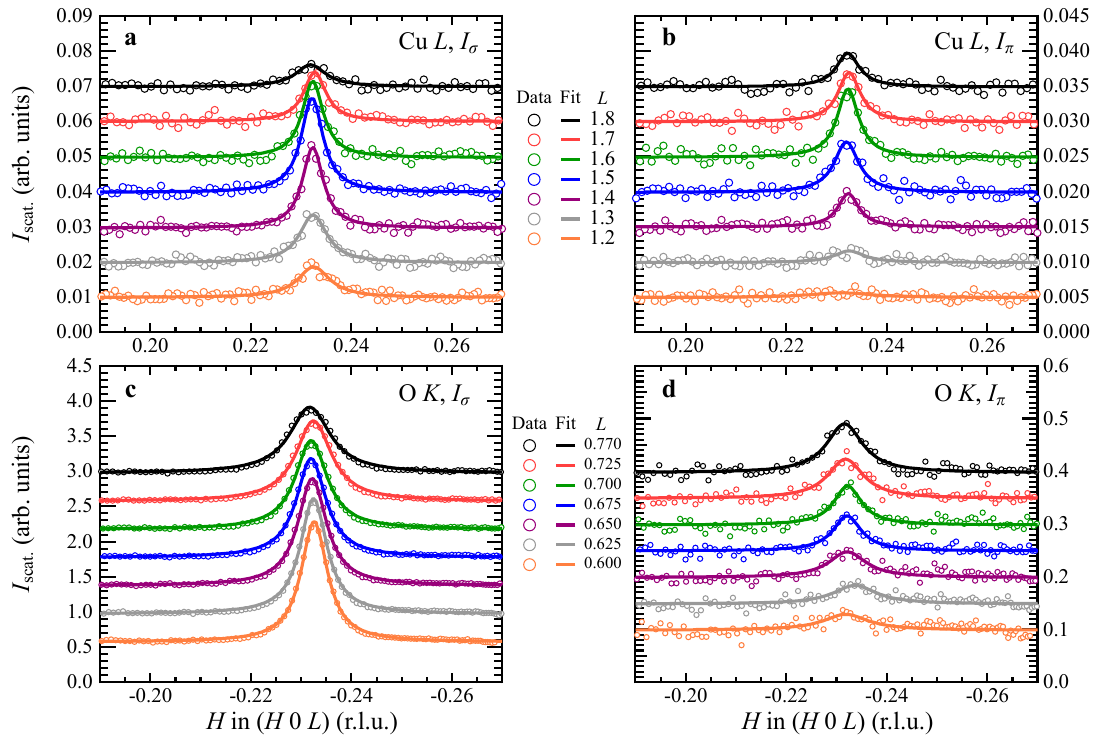}}
\caption{$L$ dependence of CDW scattering intensity in LBCO at the Cu $L$ ({\bf a}: $I_\sigma$ and {\bf b}: $I_\pi$ vs. $H$) and O $K$ ({\bf c}: $I_\sigma$ and {\bf d}: $I_\pi$ vs. $H$) edges. Open circles are background-subtracted data and solid lines are Lorenztian fits. Data are offset for clarity.}
\label{fig:Supp_LBCO_LDep}
\end{figure}

Fig.~\ref{fig:Supp_LBCO_LDep} shows the $L$ dependence of $I_\sigma$ and $I_\pi$ vs. $H$ of LBCO at the Cu $L$ edge, $\bm{Q}=(0.232~0~L)$, $\hbar\omega=931.3~\text{eV}$ (top panels) and at the O $K$ edge, $\bm{Q}=(-0.232~0~L)$, $\hbar\omega=528.3~\text{eV}$ (bottom panels). Similar to the $\phi$ dependent data, the x-ray fluorescence background was subtracted using a polynomial fit excluding the peak region. Lorentzian fits were then used to determine the scattering intensities reported in Figs.~\ref{fig:LBCO_PolDep}c and \ref{fig:LBCO_PolDep}d.

\begin{figure}[h]
\centering
\resizebox{0.8\columnwidth}{!}{\includegraphics{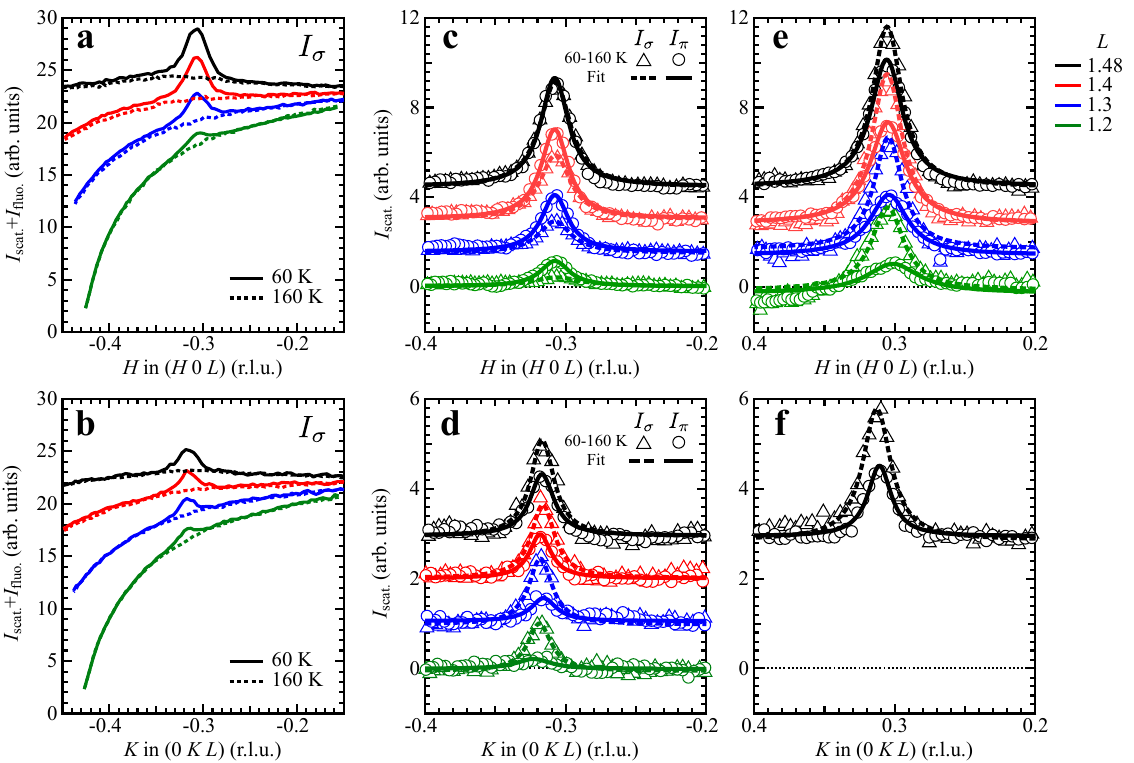}}
\caption{$L$ dependence of CDW scattering intensity in YBCO at the Cu $L$ edge for scans along $H$ (top row) and $K$ (bottom row). {\bf a-b}. Normalized intensity of scattering and x-ray fluorescence for incident $\sigma$ polarization along $H$ ({\bf a}) and $K$ ({\bf b}) at 60 K (solid lines) and 160 K (dashed lines). {\bf c-f}. Background-subtracted scattering intensity (60-160 K) for $\sigma$ (open triangles) and $\pi$ (open circles) incident photon polarization along $-H$ ({\bf c}), $-K$ ({\bf d}), $+H$ ({\bf e}) and $+K$ ({\bf f}). Dashed (solid) lines are Lorenztian fits to $I_\sigma$ ($I_\pi$). $L$ values are indicated in legend on right. Data in {\bf c-f} are offset for clarity.}
\label{fig:Supp_YBCO_LDep}
\end{figure}

Fig.~\ref{fig:Supp_YBCO_LDep} shows the $L$ dependence of $I_\sigma$ and $I_\pi$ of YBCO at the Cu $L$ edge ($\hbar\omega=931.3~\text{eV}$) for the CDW peak along $\pm H$ and $\pm K$. In this case the background subtraction was accomplished by subtracting the measured x-ray fluorescence at 160 K from the measurement at 60 K. Figs.~\ref{fig:Supp_YBCO_LDep}a and \ref{fig:Supp_YBCO_LDep}b show a representative set of such backgrounds.  Lorentzian fits were then used to determine the scattering intensities reported in Fig.~\ref{fig:YBCO_LDep}.
\\ \\ \\

\clearpage
\section{Parameter estimation and confidence regions}
The parameters $t_\parallel/t_\perp$ and $t_{cc}/t_\perp$ were determined by unconstrained, weighted least-squares fitting of the experimental data to $I(\bm{\epsilon}_{i,\pi},\omega, \bm{Q})/I(\bm{\epsilon}_{i,\sigma},\omega, \bm{Q})$, as defined by Eq. \ref{eqn:scatcorrabs}. These were the only free parameters in the fit. For LBCO at the O $K$ and Cu $L$ edges, both the $\phi$ and $L$ dependent data were fit simultaneously. This simultaneous fit narrowed the confidence regions as compared to fitting either dataset individually. Since the $\phi$ dependence at the Cu $L$ edge in YBCO was not measured here, the fits in those cases are to the $L$ dependence. Fitting to the $L$ dependence alone does not determine the signs of $t_\parallel/t_\perp$ and $t_{cc}/t_\perp$ (see Fig.~\ref{fig:Supp_ModelComparison}a), but can still constrain their absolute values and their relative sign (ie., $t_\parallel/t_\perp$ and $t_{cc}/t_\perp$ will either have the same or opposite sign). 

\begin{figure}[h]
\centering
\resizebox{0.7\columnwidth}{!}{\includegraphics{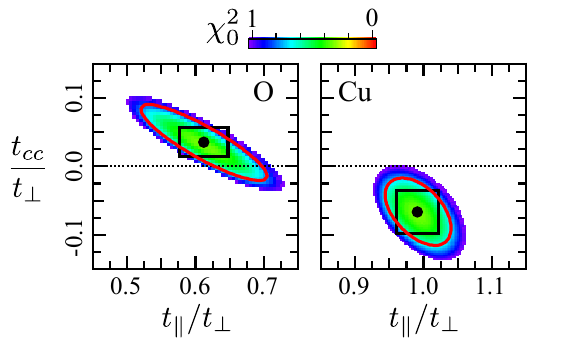}}
\caption{Maps of $\chi_0^2$ illustrating regions where model calculations capture the LBCO data recorded at the O $K$ edge (left) and Cu $L$ edge (right). The black circle is the minimum of $\chi_0^2$ and corresponds to the best fit from  nonlinear least squares fitting. The black rectangles are the 95\% confidence interval reported from the standard fitting routine, which is seen to miss a sizeable range of suitable parameters. The red ellipse is a contour of constant $\chi_0^2$ that defines a more suitable confidence {\it region} in the $t_\parallel/t_\perp$ -- $t_{cc}/t_\perp$ plane where there is a 95\% (or higher) likelihood that the experimental data is described by the model. The maps of $\chi_0^2$ are truncated at values greater than 1, emphasizing the region of good agreement.}
\label{fig:Supp_LBCO_ParamSpaceBoth}
\end{figure}

Although least-squares fitting provides standard errors for the best fit parameters, an examination of $\chi_0^2$ (the reduced chi-squared statistic) throughout the parameter space reveals that the 95\% confidence intervals defined by the standard errors underestimate the regions of high confidence. As shown in Fig.~\ref{fig:Supp_LBCO_ParamSpaceBoth}, the high confidence regions are better described by contours of constant $\chi_0^2$ in the $t_\parallel/t_\perp$ -- $t_{cc}/t_\perp$ parameter space (red ellipses in Fig.~\ref{fig:Supp_LBCO_ParamSpaceBoth}), than by the standard confidence intervals obtained by nonlinear least-squares fitting (black rectangles in Fig.~\ref{fig:Supp_LBCO_ParamSpaceBoth}). For the LBCO data at the O $K$ and Cu $L$ edges, these contours appear as ellipses in the parameter space, representing regions where there was a 95\% (or higher) likelihood that the model describes the experimental data. We note that the reported parameter uncertainties in the main text correspond to the black rectangles in Fig.~\ref{fig:Supp_LBCO_ParamSpaceBoth}, but note that the full extent of the parameters is better captured by the ellipses shown here.

\begin{figure}[h]
\centering
\resizebox{.98\columnwidth}{!}{\includegraphics{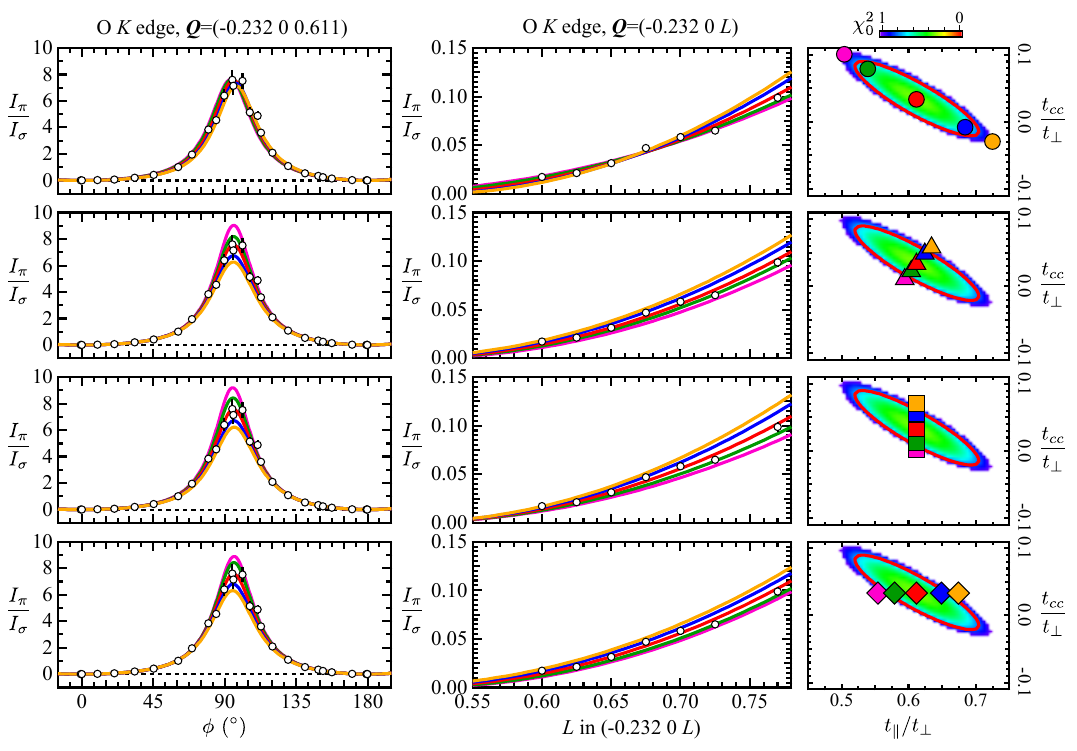}}
\caption{LBCO O $K$ edge $\phi$ (left panels) and $L$ dependence (middle panels) of measured $I_\pi/I_\sigma$ (open symbols) compared to the model calculations (coloured lines) based on Eq. \ref{eqn:scatcorrabs} at the points indicated by coloured symbols on the maps of $\chi_0^2$ in the $t_\parallel/t_\perp$ -- $t_{cc}/t_\perp$ parameter space (right panels). Each row corresponds to exploring the parameter space of the model along a particular direction (ie., diagonal, vertical, horizontal).}
\label{fig:Supp_LBCO_OK_ParamSpace}
\end{figure}

\begin{figure}[h]
\centering
\resizebox{0.8\columnwidth}{!}{\includegraphics{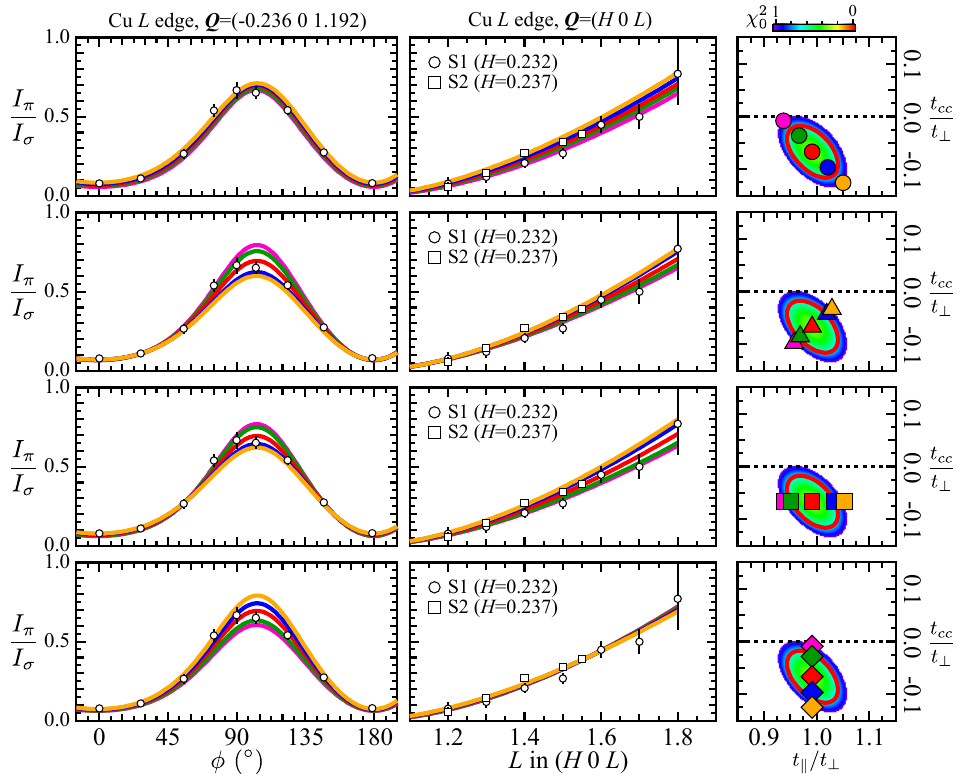}}
\caption{LBCO Cu $L$ edge $\phi$ (left panels) and $L$ dependence (middle panels) of measured $I_\pi/I_\sigma$ (open symbols) compared to the model calculations (coloured lines) based on Eq. \ref{eqn:scatcorrabs} at the points indicated by coloured symbols on the maps of $\chi_0^2$ in the $t_\parallel/t_\perp$ -- $t_{cc}/t_\perp$ parameter space (right panels). Each row corresponds to exploring the parameter space of the model along a particular direction (ie., diagonal, vertical, horizontal).}
\label{fig:Supp_LBCO_CuL_ParamSpace}
\end{figure}

\begin{figure}[ht]
\centering
\resizebox{.85\columnwidth}{!}{\includegraphics{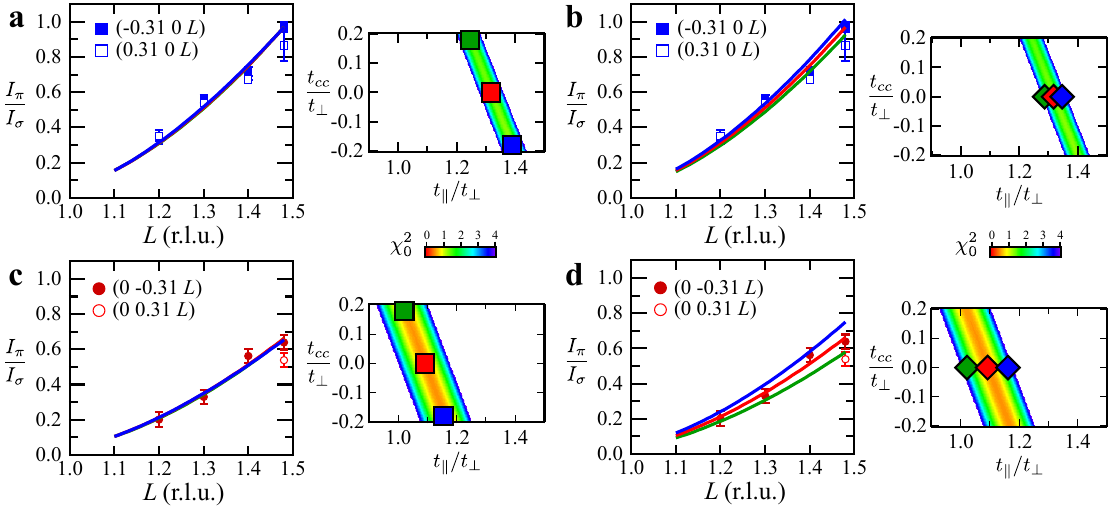}}
\caption{$L$ dependence of the model calculations based on Eq. \ref{eqn:scatcorrabs}, for YBCO at the Cu $L$ edge, at the points indicated in the colour maps of  $\chi_0^2$  in the $t_\parallel/t_\perp$ -- $t_{cc}/t_\perp$ parameter space.  The parameter space of the model is explored for the CDW peak along $H$ ({\bf a-b}) and $K$ ({\bf c-d}) in the diagonal (left panels) and horizontal (right panels) directions. The maps of $\chi_0^2$ are truncated at values greater than 4 to highlight the region of good agreement.}
\label{fig:Supp_YBCO_CuL_ParamSpace}
\end{figure}

\clearpage
To illustrate the model sensitivity in these high confidence regions, we show here how the model calculations vary across the parameter space in regions around the best fit parameters. This analysis is shown in Fig.~\ref{fig:Supp_LBCO_OK_ParamSpace} for LBCO at the O $K$ edge, Fig.~\ref{fig:Supp_LBCO_CuL_ParamSpace} for LBCO at the Cu $L$ edge, and Fig.~\ref{fig:Supp_YBCO_CuL_ParamSpace} for YBCO at the Cu $L$ edge. 

We note that the absolute values of $\chi_0^2$ reported here should not be over-interpreted as the formal definition of $\chi_0^2$ requires that weights be calculated from true variances (ie. $\sigma$ from a normal distribution of repeated measurements). Here, we used statistical errors from fitting the CDW peak to estimate the variance, which can lead to $\chi_0^2<1$, indicating that the experimental uncertainty underestimates the variance. Although the magnitudes of   $\chi_0^2$ are affected by this detail, one can still use this statistic as a means of identifying regions of parameter space where the model yields good agreement with the data. It is in this sense that we have opted to define the high confidence regions of parameter space where the model agrees with the data. 

\clearpage

\begin{figure}[h]
\centering
\resizebox{0.9\columnwidth}{!}{\includegraphics{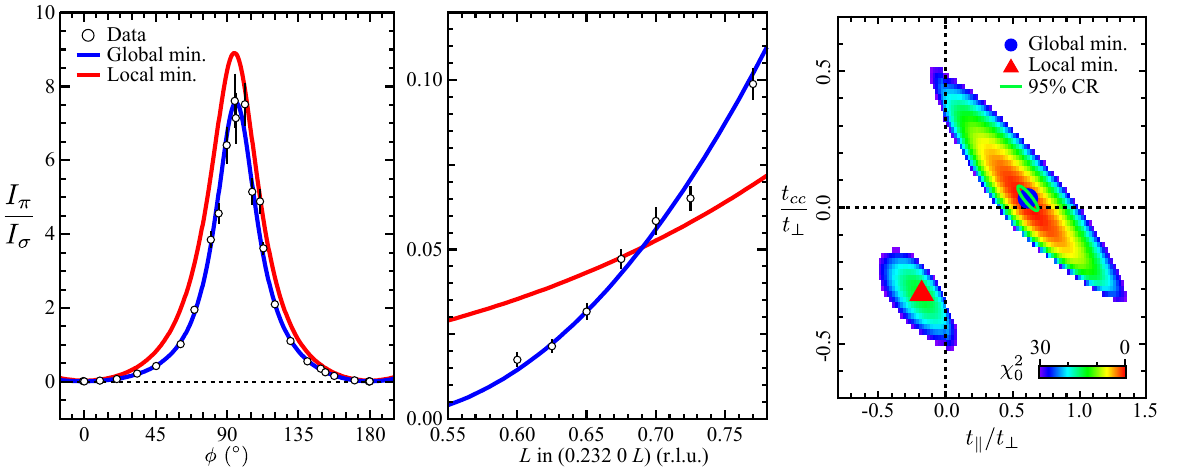}}
\caption{Comparison of global minimum in $\chi_0^2$ with $t_\parallel/t_\perp>0$ with local minimum having $t_\parallel/t_\perp<0$ for LBCO at the O $K$ edge. Model calculations plotted against $\phi$ (left panel) and $L$ (center panel) for the global and local minima indicated by coloured symbols in the right panel. The colour map of $\chi_0^2$ is truncated above $\chi_0^2=30$. The $95\%$ confidence region , shown as a light green ellipse, is the same as in Figs.~\ref{fig:LBCO_PolDep}e and \ref{fig:Supp_LBCO_OK_ParamSpace}. The global minimum matches the data, whereas the local minimum with $t_\parallel/t_\perp<0$ (corresponding to $|\Delta_d|>|\Delta_{s'}|$) is a poor match. }
\label{fig:Supp_LBCO_OKL_ParamSpace_posNeg}
\end{figure}

Lastly, we illustrate the confidence with which the sign of $t_\parallel/t_\perp$ is reported for LBCO at the O $K$ edge. An examination of $\chi_0^2$ over a larger region of the parameter space is shown in right panel of Fig.~\ref{fig:Supp_LBCO_OKL_ParamSpace_posNeg}. Here $\chi_0^2$ is truncated above $\chi_0^2=30$. The value of $\chi_0^2$ outside the plotted region was found to be very large. The same $95\%$ confidence region as shown above (Fig.~\ref{fig:Supp_LBCO_ParamSpaceBoth}, left) is shown in light green to provide a sense of scale for the parameter space. In addition to the global minimum at $t_{\parallel}/ t_{\perp} = 0.612$ and $t_{cc}/t_{\perp} = 0.034$, we find a local minimum in $\chi_0^2$ with a small and negative $t_\parallel/t_\perp$ and a relatively large and negative $t_{cc}/t_\perp$. As shown in Fig.~\ref{fig:Supp_LBCO_OKL_ParamSpace_posNeg}, this local minimum has a considerably higher $\chi_0^2$ than the global minimum (right panel) and is a poor fit to the data along $\phi$ (left panel) and $L$ (center panel). A similar examination of $\chi_0^2$ throughout the parameter space for the Cu $L$ edge data in LBCO reveals very high values of  $\chi_0^2$ for $t_\parallel/t_\perp<0$ and plots of the model for these values are in very poor agreement with the measured $\phi$ dependence (not shown). We therefore have a high degree of confidence that $t_\parallel/t_\perp>0$,  ruling out a dominant $d$ symmetry to the CDW order in LBCO within the theoretical framework presented here.

\bibliographystyle{nphys_mod}

\begin{thebibliography}{10}

\bibitem{Tranquada95}
{Tranquada}, J.~M., {Sternlieb}, B.~J., {Axe}, J.~D., {Nakamura}, Y. \&
  {Uchida}, S., Evidence for stripe correlations of spins and holes in copper
  oxide superconductors. \textit{Nature} \textbf{375}, 561--563 (1995).

\bibitem{Kohsaka07}
Kohsaka, Y. \textit{et~al.}, An Intrinsic Bond-Centered Electronic Glass with
  Unidirectional Domains in Underdoped Cuprates. \textit{Science} \textbf{315},
  5817, 1380--1385 (2007).

\bibitem{Wu11}
{Wu}, T. \textit{et~al.}, {Magnetic-field-induced charge-stripe order in the
  high-temperature superconductor YBa$_{2}$Cu$_{3}$O$_{y}$}. \textit{Nature}
  \textbf{477}, 191--194 (2011).

\bibitem{Ghiringhelli12}
Ghiringhelli, G. \textit{et~al.}, Long-Range Incommensurate Charge Fluctuations
  in {(Y,Nd)Ba$_2$Cu$_3$O$_{6+x}$}. \textit{Science} \textbf{337}, 6096,
  821--825 (2012).

\bibitem{Chang12}
{Chang}, J. \textit{et~al.}, {Direct observation of competition between
  superconductivity and charge density wave order in
  YBa$_{2}$Cu$_{3}$O$_{6.67}$}. \textit{Nature Phys.} \textbf{8}, 871--876
  (2012).

\bibitem{Achkar12}
{Achkar}, A.~J. \textit{et~al.}, {Distinct Charge Orders in the Planes and
  Chains of Ortho-III-Ordered YBa$_{2}$Cu$_{3}$O$_{6+\delta}$ Superconductors
  Identified by Resonant Elastic X-ray Scattering}. \textit{Phys. Rev. Lett.}
  \textbf{109}, 16, 167001 (2012).

\bibitem{Blackburn13}
Blackburn, E. \textit{et~al.}, {X-Ray Diffraction Observations of a
  Charge-Density-Wave Order in Superconducting Ortho-II
  YBa$_{2}$Cu$_{3}$O$_{6.54}$ Single Crystals in Zero Magnetic Field}.
  \textit{Phys. Rev. Lett.} \textbf{110}, 137004 (2013).

\bibitem{Comin14a}
Comin, R. \textit{et~al.}, Charge Order Driven by Fermi-Arc Instability in
  Bi$_2$Sr$_{2-x}$La$_x$CuO$_{6+\delta}$. \textit{Science} \textbf{343}, 6169,
  390--392 (2014).

\bibitem{daSilvaNeto14}
da~Silva~Neto, E.~H. \textit{et~al.}, Ubiquitous Interplay Between Charge
  Ordering and High-Temperature Superconductivity in Cuprates. \textit{Science}
  \textbf{343}, 6169, 393--396 (2014).

\bibitem{Vojta09}
{Vojta}, M., {Lattice symmetry breaking in cuprate superconductors: stripes,
  nematics, and superconductivity}. \textit{Adv. Phys.} \textbf{58}, 699--820
  (2009).

\bibitem{Lawler10}
{Lawler}, M.~J. \textit{et~al.}, {Intra-unit-cell electronic nematicity of the
  high-T$_{c}$ copper-oxide pseudogap states}. \textit{Nature} \textbf{466},
  347--351 (2010).

\bibitem{Li06b}
Li, J.-X., Wu, C.-Q. \& Lee, D.-H., Checkerboard charge density wave and
  pseudogap of high-${T}_{c}$ cuprate. \textit{Phys. Rev. B} \textbf{74},
  184515 (2006).

\bibitem{Seo07}
Seo, K., Chen, H.-D. \& Hu, J., $d$-wave checkerboard order in cuprates.
  \textit{Phys. Rev. B} \textbf{76}, 020511 (2007).

\bibitem{Vojta08}
Vojta, M. \& R\"osch, O., Superconducting $d$-wave stripes in cuprates: Valence
  bond order coexisting with nodal quasiparticles. \textit{Phys. Rev. B}
  \textbf{77}, 094504 (2008).

\bibitem{Metlitski10}
Metlitski, M.~A. \& Sachdev, S., Quantum phase transitions of metals in two
  spatial dimensions. II. Spin density wave order. \textit{Phys. Rev. B}
  \textbf{82}, 075128 (2010).

\bibitem{Sachdev13}
Sachdev, S. \& La~Placa, R., Bond Order in Two-Dimensional Metals with
  Antiferromagnetic Exchange Interactions. \textit{Phys. Rev. Lett.}
  \textbf{111}, 027202 (2013).

\bibitem{Efetov13}
{Efetov}, K.~B., {Meier}, H. \& {P{\'e}pin}, C., {Pseudogap state near a
  quantum critical point}. \textit{Nature Phys.} \textbf{9}, 442--446 (2013).

\bibitem{Atkinson14}
Atkinson, W.~A., Kampf, A.~P. \& Bulut, S., Charge order in the pseudogap phase
  of cuprate superconductors. \textit{New Journal of Physics} \textbf{17}, 1,
  013025 (2015).

\bibitem{Fujita14}
Fujita, K. \textit{et~al.}, Direct phase-sensitive identification of a $d$-form
  factor density wave in underdoped cuprates. \textit{Proc. Natl. Acad. Sci.
  U.S.A.} \textbf{111}, 30, E3026--E3032 (2014).

\bibitem{Allais14}
Allais, A., Bauer, J. \& Sachdev, S., Density wave instabilities in a
  correlated two-dimensional metal. \textit{Phys. Rev. B} \textbf{90}, 155114
  (2014).

\bibitem{Chowdhury14b}
Chowdhury, D. \& Sachdev, S., Density-wave instabilities of fractionalized
  Fermi liquids. \textit{Phys. Rev. B} \textbf{90}, 245136 (2014).

\bibitem{Comin14c}
Comin, R. \textit{et~al.}, {Symmetry of charge order in cuprates}.
  \textit{Nature Materials} \textbf{14}, 796 (2015).

\bibitem{Hucker11}
H\"ucker, M. \textit{et~al.}, Stripe order in superconducting
  {La$_{2-x}$Ba$_x$CuO$_4$} ($0.095\leqslant x\leqslant 0.155$). \textit{Phys.
  Rev. B} \textbf{83}, 104506 (2011).

\bibitem{Fujita14a}
Fujita, K. \textit{et~al.}, Simultaneous Transitions in Cuprate Momentum-Space
  Topology and Electronic Symmetry Breaking. \textit{Science} \textbf{344},
  6184, 612--616 (2014).

\bibitem{Huecker14}
H\"ucker, M. \textit{et~al.}, Competing charge, spin, and superconducting
  orders in underdoped {YBa$_{2}$Cu$_{3}$O$_{y}$}. \textit{Phys. Rev. B}
  \textbf{90}, 054514 (2014).

\bibitem{BlancoCanosa14}
Blanco-Canosa, S. \textit{et~al.}, Resonant x-ray scattering study of
  charge-density wave correlations in {YBa$_2$Cu$_3$O$_{6+x}$}. \textit{Phys.
  Rev. B} \textbf{90}, 054513 (2014).

\bibitem{Achkar13}
Achkar, A.~J. \textit{et~al.}, Resonant X-Ray Scattering Measurements of a
  Spatial Modulation of the Cu $3d$ and O $2p$ Energies in Stripe-Ordered
  Cuprate Superconductors. \textit{Phys. Rev. Lett.} \textbf{110}, 017001
  (2013).

\bibitem{Fujita12}
{Fujita}, K. \textit{et~al.}, {Spectroscopic Imaging Scanning Tunneling
  Microscopy Studies of Electronic Structure in the Superconducting and
  Pseudogap Phases of Cuprate High-{$T_\mathrm{c}$} Superconductors}.
  \textit{J. Phys. Soc. Jpn.} \textbf{81}, 1, 011005 (2012).

\bibitem{Wise08}
{Wise}, W.~D. \textit{et~al.}, {Charge-density-wave origin of cuprate
  checkerboard visualized by scanning tunnelling microscopy}. \textit{Nature
  Phys.} \textbf{4}, 696 (2008).

\bibitem{Yamada98}
Yamada, K. \textit{et~al.}, {Doping dependence of the spatially modulated
  dynamical spin correlations and the superconducting-transition temperature in
  La$_{2-x}$Sr$_{x}$CuO$_{4}$}. \textit{Phys. Rev. B} \textbf{57}, 10,
  6165--6172 (1998).

\bibitem{Kampf01}
Kampf, A.~P., Scalapino, D.~J. \& White, S.~R., Stripe orientation in an
  anisotropic $t-J$ model. \textit{Phys. Rev. B} \textbf{64}, 052509 (2001).

\bibitem{Blanco-Canosa13}
Blanco-Canosa, S. \textit{et~al.}, Momentum-Dependent Charge Correlations in
  {YBa$_2$Cu$_3$O$_{6+\delta}$} Superconductors Probed by Resonant X-Ray
  Scattering: Evidence for Three Competing Phases. \textit{Phys. Rev. Lett.}
  \textbf{110}, 187001 (2013).

\bibitem{Comin15}
Comin, R. \textit{et~al.}, Broken translational and rotational symmetry via
  charge stripe order in underdoped {YBa$_2$Cu$_3$O$_{6+y}$}. \textit{Science}
  \textbf{347}, 6228, 1335--1339 (2015).

\bibitem{Abbamonte05}
Abbamonte, P. \textit{et~al.}, Spatially modulated `{M}ottness' in
  {La$_{2-x}$Ba$_x$CuO$_4$}. \textit{Nat. Phys.} \textbf{1}, 155--158 (2005).

\bibitem{Fink09}
Fink, J. \textit{et~al.}, Charge ordering in {La$_{1.8 -
  x}$Eu$_{0.2}$Sr$_x$CuO$_4$} studied by resonant soft x-ray diffraction.
  \textit{Phys. Rev. B} \textbf{79}, 10, 100502 (2009).

\bibitem{Wu12}
{Wu}, H.-H. \textit{et~al.}, {Charge stripe order near the surface of
  12-percent doped La$_{2-x}$Sr$_{x}$CuO$_{4}$}. \textit{Nature Commun.}
  \textbf{3}, 1023 (2012).

\bibitem{Benjamin13}
Benjamin, D., Abanin, D., Abbamonte, P. \& Demler, E., {Microscopic Theory of
  Resonant Soft-X-Ray Scattering in Materials with Charge Order: The Example of
  Charge Stripes in High-Temperature Cuprate Superconductors}. \textit{Phys.
  Rev. Lett.} \textbf{110}, 13, 137002 (2013).

\bibitem{Sachdev03}
Sachdev, S., \textit{Colloquium:} \quad{}Order and quantum phase transitions in
  the cuprate superconductors. \textit{Rev. Mod. Phys.} \textbf{75}, 3,
  913--932 (2003).

\bibitem{Christensen07}
Christensen, N.~B. \textit{et~al.}, Nature of the Magnetic Order in the
  Charge-Ordered Cuprate {La$_{1.48}$Nd$_{0.4}$Sr$_{0.12}$CuO$_{4}$}.
  \textit{Phys. Rev. Lett.} \textbf{98}, 197003 (2007).

\bibitem{Fischer14}
Fischer, M.~H., Wu, S., Lawler, M., Paramekanti, A. \& Kim, E.-A., Nematic and
  spin-charge orders driven by hole-doping a charge-transfer insulator.
  \textit{New Journal of Physics} \textbf{16}, 9, 093057 (2014).

\bibitem{Yamakawa14}
Yamakawa, Y. \& Kontani, H., Spin-Fluctuation-Driven Nematic Charge-Density
  Wave in Cuprate Superconductors: Impact of Aslamazov-Larkin Vertex
  Corrections. \textit{Phys. Rev. Lett.} \textbf{114}, 257001 (2015).

\bibitem{Chowdhury14}
Chowdhury, D. \& Sachdev, S., Feedback of superconducting fluctuations on
  charge order in the underdoped cuprates. \textit{Phys. Rev. B} \textbf{90},
  134516 (2014).

\bibitem{Lee14}
Lee, P.~A., Amperean Pairing and the Pseudogap Phase of Cuprate
  Superconductors. \textit{Phys. Rev. X} \textbf{4}, 031017 (2014).

\bibitem{Meier14}
Meier, H., P\'epin, C., Einenkel, M. \& Efetov, K.~B., Cascade of phase
  transitions in the vicinity of a quantum critical point. \textit{Phys. Rev.
  B} \textbf{89}, 195115 (2014).

\bibitem{Pepin14}
P\'epin, C., de~Carvalho, V.~S., Kloss, T. \& Montiel, X., Pseudogap, charge
  order, and pairing density wave at the hot spots in cuprate superconductors.
  \textit{Phys. Rev. B} \textbf{90}, 195207 (2014).

\bibitem{Wang14}
Wang, Y. \& Chubukov, A., Charge-density-wave order with momentum $(2Q,0)$ and
  $(0,2Q)$ within the spin-fermion model: Continuous and discrete symmetry
  breaking, preemptive composite order, and relation to pseudogap in hole-doped
  cuprates. \textit{Phys. Rev. B} \textbf{90}, 035149 (2014).

\bibitem{Hawthorn11a}
Hawthorn, D.~G. \textit{et~al.}, An in-vacuum diffractometer for resonant
  elastic soft x-ray scattering. \textit{Rev. Sci. Instrum.} \textbf{82}, 7,
  073104 (2011).

\bibitem{Grafe10}
Grafe, H.-J. \textit{et~al.}, Charge order and low frequency spin dynamics in
  lanthanum cuprates revealed by Nuclear Magnetic Resonance. \textit{The
  European Physical Journal Special Topics} \textbf{188}, 1, 89--101 (2010).

\end{thebibliography}

\begin{thebibliography}{10}

\bibitem{Comin14cs}
Comin, R. \textit{et~al.}, {Symmetry of charge order in cuprates}.
  \textit{Nature Materials} \textbf{14}, 796 (2015).

\bibitem{Abbamonte05s}
Abbamonte, P. \textit{et~al.}, Spatially modulated `{M}ottness' in
  {La$_{2-x}$Ba$_x$CuO$_4$}. \textit{Nat. Phys.} \textbf{1}, 155--158 (2005).

\bibitem{Achkar14as}
Achkar, A.~J. \textit{et~al.}, {Impact of Quenched Oxygen Disorder on Charge
  Density Wave Order in {YBa$_2$Cu$_3$O$_{6+x}$}}. \textit{Phys. Rev. Lett.}
  \textbf{113}, 107002 (2014).

\bibitem{Chen92s}
Chen, C. \textit{et~al.}, Out-of-plane orbital characters of intrinsic and
  doped holes in {La$_{2-x}$Sr$_x$CuO$_4$}. \textit{Phys. Rev. Lett.}
  \textbf{68}, 16, 2543--2546 (1992).

\bibitem{Hawthorn11bs}
Hawthorn, D.~G. \textit{et~al.}, Resonant elastic soft x-ray scattering in
  oxygen-ordered {YBa$_2$Cu$_3$O$_{6+\delta}$}. \textit{Phys. Rev. B}
  \textbf{84}, 075125 (2011).

\bibitem{Achkar13s}
Achkar, A.~J. \textit{et~al.}, Resonant X-Ray Scattering Measurements of a
  Spatial Modulation of the Cu $3d$ and O $2p$ Energies in Stripe-Ordered
  Cuprate Superconductors. \textit{Phys. Rev. Lett.} \textbf{110}, 017001
  (2013).

\bibitem{Sachdev13s}
Sachdev, S. \& La~Placa, R., Bond Order in Two-Dimensional Metals with
  Antiferromagnetic Exchange Interactions. \textit{Phys. Rev. Lett.}
  \textbf{111}, 027202 (2013).

\bibitem{Achkar12s}
{Achkar}, A.~J. \textit{et~al.}, {Distinct Charge Orders in the Planes and
  Chains of Ortho-III-Ordered YBa$_{2}$Cu$_{3}$O$_{6+\delta}$ Superconductors
  Identified by Resonant Elastic X-ray Scattering}. \textit{Phys. Rev. Lett.}
  \textbf{109}, 16, 167001 (2012).

\bibitem{Fischer14s}
Fischer, M.~H., Wu, S., Lawler, M., Paramekanti, A. \& Kim, E.-A., Nematic and
  spin-charge orders driven by hole-doping a charge-transfer insulator.
  \textit{New Journal of Physics} \textbf{16}, 9, 093057 (2014).

\bibitem{Thomson14s}
Thomson, A. \& Sachdev, S., Charge ordering in three-band models of the
  cuprates. \textit{Phys. Rev. B} \textbf{91}, 115142 (2015).

\bibitem{Kim08s}
Kim, Y.-J., Gu, G.~D., Gog, T. \& Casa, D., X-ray scattering study of charge
  density waves in {La$_{2-x}$Ba$_x$CuO$_4$}. \textit{Phys. Rev. B}
  \textbf{77}, 6, 064520 (2008).

\end{thebibliography}

\end{document}